\documentclass[iop, twocolumn]{aastex63}
\hypersetup{citecolor=blue,urlcolor=red}
\usepackage{multirow}
\usepackage{gensymb}
\usepackage{amsmath}
\usepackage{graphicx}
\usepackage{float}
\usepackage{subfigure}
\usepackage{lineno}
      
\shorttitle{PSRs J0941$-$39 and J1107$-$5907}
\shortauthors{Sun et al.}

\begin{document}

\title{Wide Bandwidth Observations of PSR J0941$-$39 and PSR J1107$-$5907}

\correspondingauthor{N. Wang}
\email{na.wang@xao.ac.cn}

\author{S. N. Sun}
\affiliation{Xinjiang Astronomical Observatory, Chinese Academy of Sciences, Urumqi, Xinjiang 830011, China}
\affiliation{Xinjiang Key Laboratory of Radio Astrophysics, 150 Science1-Street, Urumqi 830011, China}

\author{N. Wang}
\affiliation{Xinjiang Astronomical Observatory, Chinese Academy of Sciences, Urumqi, Xinjiang 830011, China}
\affiliation{Xinjiang Key Laboratory of Radio Astrophysics, 150 Science1-Street, Urumqi 830011, China}
\affiliation{Key Laboratory of Radio Astronomy, Chinese Academy of Sciences, Nanjing 210008, China}

\author{W. M. Yan\href{https://orcid.org/0000-0002-7662-3875}}
\affiliation{Xinjiang Astronomical Observatory, Chinese Academy of Sciences, Urumqi, Xinjiang 830011, China}
\affiliation{Xinjiang Key Laboratory of Radio Astrophysics, 150 Science1-Street, Urumqi 830011, China}
\affiliation{Key Laboratory of Radio Astronomy, Chinese Academy of Sciences, Nanjing 210008, China}

\author{S. Q. Wang}
\affiliation{Xinjiang Astronomical Observatory, Chinese Academy of Sciences, Urumqi, Xinjiang 830011, China}
\affiliation{Xinjiang Key Laboratory of Radio Astrophysics, 150 Science1-Street, Urumqi 830011, China}

\author{J. T. Xie}
\affiliation{Research Center for Intelligent Computing Platforms, Zhejiang Laboratory, Hangzhou 311100, China}

\begin{abstract}

We present a polarization analysis of PSR J0941$-$39 and PSR J1107$-$5907, which exhibit transitions between being pulsars and rotating radio transients (RRATs), {using the ultra-wide bandwidth low-frequency (UWL) receiver} on Murriyang, the Parkes 64\,m radio telescope. The spectral index of each pulsar was measured, revealing distinct variations among different states. By using the rotating vector model (RVM), we determined that the magnetosphere geometry remains consistent between the RRAT state and the pulsar state for PSR J0941$-$39, with emissions originating from the same height in the magnetosphere. The occurrence of the RRAT state could be attributed to variations in currents within the pulsar's magnetosphere. Our results suggest that the emission mechanism of RRAT may share similarities with that of a typical pulsar.

\end{abstract}


\section{INTRODUCTION}

Rotating radio transients (RRATs) are a subclass of pulsars that were discovered through a single pulse search of archival data from the Parkes multi-beam pulsar survey~\citep{mll+2006}. These pulsars exhibit bright bursts of radio emission with durations in the millisecond range, occurring sporadically every few minutes to hours~\citep{kkl+2011}. It is believed that RRATs are neutron stars with highly variable emissions~\citep{kkl+2011,zhx+2023}. To date, more than 100 RRATs have been detected. Similar to normal pulsars, the pulse energies of RRATs follow log-normal distributions~\citep{cbm+2017,mmm+2018}. However, RRATs exhibit higher magnetic fields compared to normal pulsars~\citep{h2013}. The origin of RRAT emissions remains unclear. Numerous theories have been proposed to explain the peculiar emission properties exhibited by RRATs. These include disruptions in radio emission caused by contamination of the pulsar magnetosphere due to fallback supernova material~\citep{l2006}, radiation belts surrounding the star~\citep{lm2007}, or debris present in its vicinity~\citep{cs2008}.

\begin{table*}
\renewcommand\arraystretch{1.3}
\centering
\caption{Observations results of two states for PSR J0941$-$39 and PSR J1107$-$5907.}
\label{obs2}
\begin{tabular}{ccccccccc}
\hline
NAME & MJD  &  Total pulse number &   State    & Continuous number  & RM & RM$_{\rm ion}$ & RM$_{\rm ISM}$  \\
           &           &             &                                    &                                   & ${\rm (rad\,m^{-2}})$ & ${\rm (rad\,m^{-2}})$ & ${\rm (rad\,m^{-2}})$ \\     
\hline
\multirow{3}{*} {J0941$-$39}   &   59598.56  &  8228    &  RRAT     &       0-8227          &  $-74.2(1)$  & $0.72 (9)$ &  $-74.9(2)$  \\                          
                          &  59744.31  &   2534   &   Pulsar  &       0-2533        &  $-76.02 (4)$  & $3.4(2)$ &  $-79.4(2)$ \\
                          &  59897.73  &  11574  &   RRAT    &        0-11573       &    $ -74.0 (1) $  &  $1.6(1)$ &  $-75.7(2)$
\\  
\multirow{6}{*}{J1107$-$5907}  &  \multirow{6}{*}{58653.09}  & \multirow{6}{*}{134580}  &  \multirow{3}{*}{RRAT}     &  0-6434  &    & \\
&               &                      &                   &  8650-49035    &       &    & \\
&               &                      &                   &  55470-134579  &   \multirow{2}{*}{ $23.99(5)$}  & \multirow{2}{*}{$1.40(8)$} & \multirow{2}{*}{ $22.6(1)$} \\     
&               &     &\multirow{2}{*}{Pulsar}  &  6435-8649 &                          & \\
&               &      &                                      &49036-55469 &                      &    & \\                         
\hline
 \end{tabular}
\end{table*}

The discovery of pulsars, which exhibit transitions between persistent emissions (pulsar state) and sporadic emissions (RRAT state), appears to establish a connection between pulsar nulling and RRATs~\citep{bb2010}.
Four such pulsars have been detected, namely PSR B0826$-$34~\citep{eam+2012}, PSR J0941$-$39~\citep{bb2010}, PSR J1107$-$5907~\citep{yws+2014}, and PSR J1752+2359~\citep{gjw2014,syw+2021}. It remains uncertain whether nulling pulsars can undergo an evolutionary transition to become RRATs. Through the analysis of a substantial sample of nulling pulsars, it was discovered that there is a potential increase in the fraction of {nulling with the age}~\citep{wmj2007}. In theory, with {the increasing age}, sporadic maintenance of pair production processes in {their pulsar magnetospheres} could lead to the emergence of RRAT phenomena~\citep{zgd+2007}. {All four pulsars are significantly older} compared to normal pulsars, with characteristic ages on the order of $\sim10^8$\,yr~\citep{lfl+2006,Reser2013}. This further supports the hypothesis that RRATs may represent an extreme manifestation of nulling pulsars. However, it is still unknown whether the emission mechanism of RRATs is the same as that of normal pulsars.

The polarimetric observation of this subclass of pulsars can provide further insights into the emission characteristics of RRATs. In this study, we used {the ultra-wide bandwidth low-frequency (UWL) receiver} on the Parkes 64\,m radio telescope, which offers continuous frequency coverage ranging from 704\,MHz to 4032\,MHz~\citep{hmd+2020}, to carry out observations on PSR J0941$-$39 and PSR J1107$-$5907.

The discovery of PSR J0941$-$39 was made in archival surveys of pulsars at intermediate Galactic latitudes~\citep{bb2010}. The characteristic age of this pulsar is estimated to be $1.89\times10^{8}$\,yr~\citep{Reser2013}. During its pulsar state, the pulsar exhibits complicated subpulse drifting phenomenon, while sporadic bursts occur during its RRAT state with a burst rate ranging from approximately 90 to 110 per hour~\citep{bb2010}. The duration of the RRAT/pulsar state can last for several hours or even longer.

PSR J1107$-$5907, which was discovered in the Parkes 20-cm multibeam pulsar survey of the Galactic plane~\citep{lfl+2006}, is also classified as an aged pulsar with a characteristic age of $4.45\times10^{8}$\,yr. It exhibits three distinct emission states: an off state characterized by no detectable emission, a weak state, and a strong state~\citep{okl+2006, whk+2020}. \citet{yws+2014} proposed that the off state may represent an extreme manifestation of the weak state. The duration of each emission state for this pulsar ranges from several minutes to several hours~\citep{hhb+2016}. \citet{mtb+2018} presented simultaneous observations of PSR J1107-5907 with the Murchison Widefield Array (MWA) at 154 MHz and the recently upgraded Molonglo Observatory Synthesis Telescope (UTMOST) at 835 MHz, and they found that pulse energy distribution of the bright state follows a log-normal distribution at both frequencies, rather than a power law distribution.

In this paper, the observations are described in Section~\ref{sec:obs}, the results are presented in Section~\ref{sec:res}, the discussion and summary are provided in Section~\ref{sec:discussion}.

\begin{figure*}
\centering
 \begin{minipage}[t]{0.45\textwidth}
 \includegraphics[width=\columnwidth]{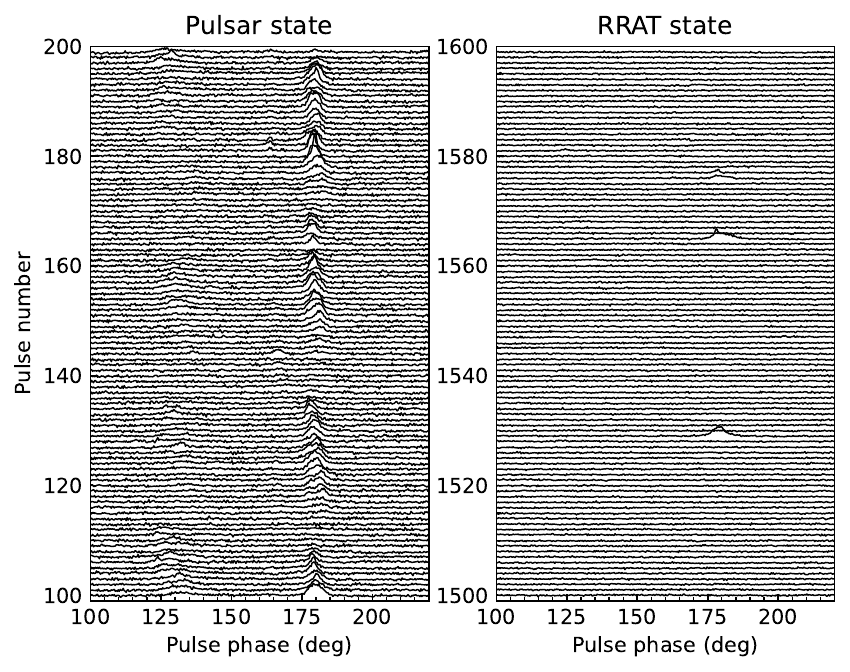}
\centerline{(a) PSR J0941$-$39}
\end{minipage}
 \begin{minipage}[t]{0.45\textwidth}
\includegraphics[width=\columnwidth]{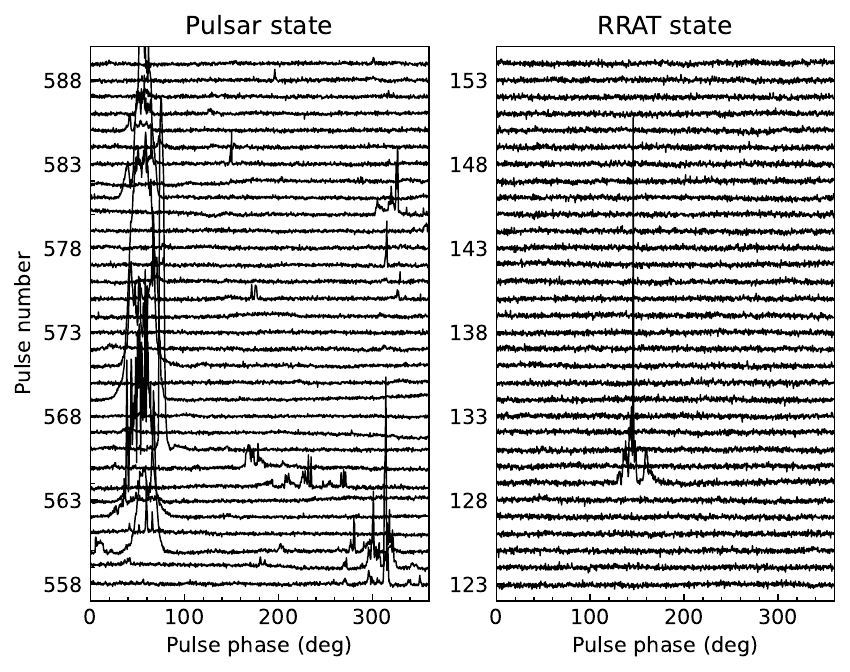}
 \centerline{(b) PSR J1107$-$5907}
  \end{minipage}
  \caption{Single-pulse stacks of PSR J0941$-$39 (panel (a)) and PSR J1107$-$5907 (panel (b)) during pulsar state and RRAT state.}
 \label{fig:1state}
 \end{figure*}

 \begin{figure*}
\centering
\begin{minipage}{0.45\textwidth}
\centerline{\includegraphics[width=\columnwidth]{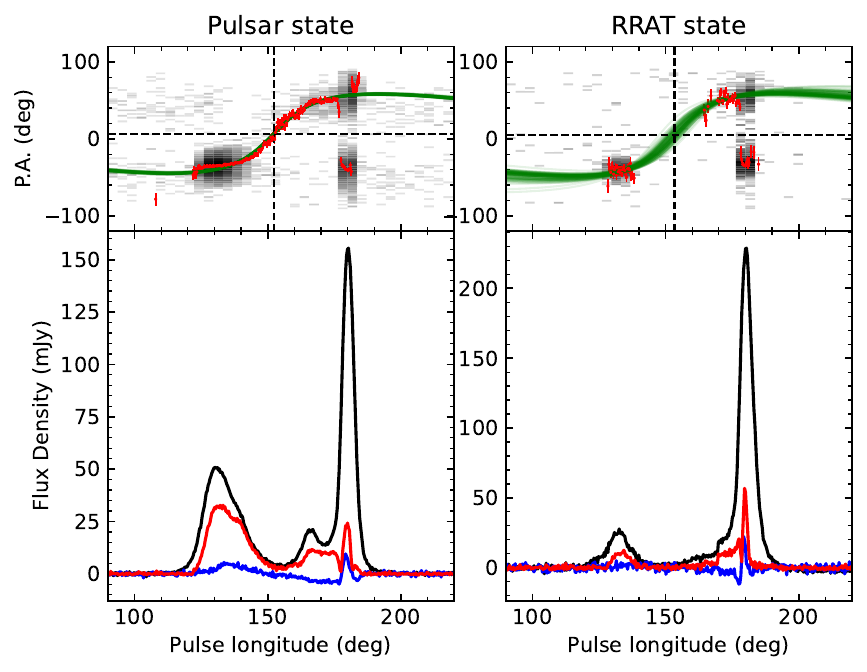}}
\centerline{(a) PSR J0941$-$39}
\end{minipage}
\begin{minipage}{0.45\textwidth}
\centerline{\includegraphics[width=\columnwidth]{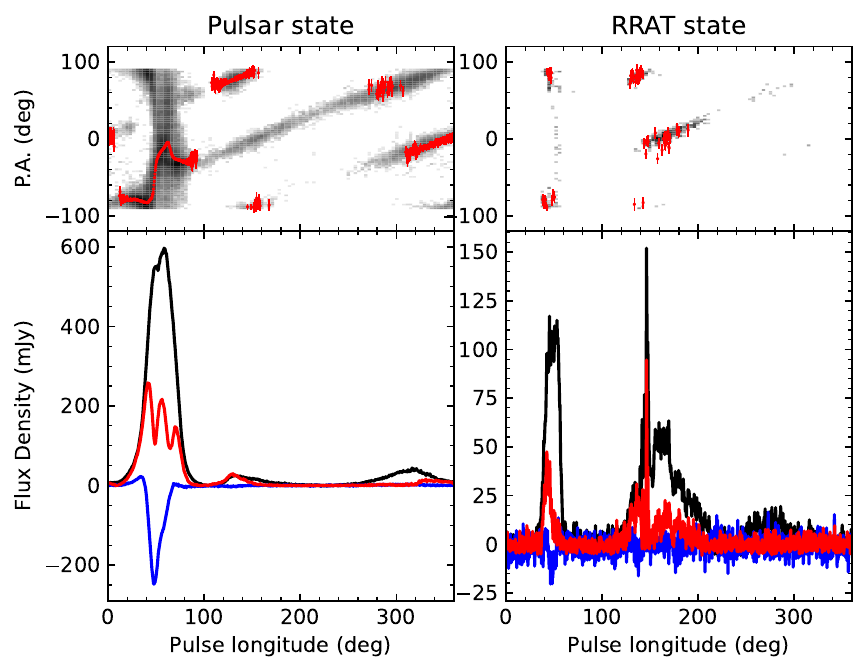}}
\centerline{(b) PSR J1107$-$5907}
\end{minipage}
 \caption{Polarization profiles for pulsar state and RRAT state for PSR J0941$-$39 (panel (a)) and PSR J1107$-$5907 (panel (b)). The black, red, and blue lines are for the total, linear, and circular polarizations, respectively. The red and gray dots in the upper panels are for the PAs of the average profile and single pulses, respectively. The green lines are for the RVM fitting results.}
 \label{fig:1pa}
 \end{figure*}

\section{OBSERVATIONS AND DATA PROCESSING}\label{sec:obs}

We carried out observations of PSR J0941$-$39 and PSR J1107$-$5907 using the UWL receiver on Murriyang, the Parkes 64\,m radio telescope, with the Medusa backend. For our observations, we recorded search-mode data with 8-bit sampling, 4 polarizations, 3328 channels for the whole band, and a sampling interval of $128\,\mu s$. For PSR J0941$-$39,  observations were carried out on January 19, June 14, and November 14, 2022. For PSR J1107$-$5907, the data was obtained from the Parkes Pulsar Data Archive\footnote{\url{https://data.csiro.au}}~\citep{hmm+2011}, which was observed on June 19, 2019.

In the data analysis, we used the {\tt DSPSR}~\citep{vb2011} software package to fold search-mode data. At each edge of the 26 sub-bands, we removed 5\,MHz of the bandpass. Subsequently, we used {\tt paz} and {\tt pazi} in the {\tt PSRCHIVE} software package~\citep{hvm2004} to remove the radio frequency interference (RFI) of the data. Calibration file where a noise diode signal was injected into the feed before each observation was recorded, and then the data was calibrated using {\tt pac} in {\tt PSRCHIVE} to transform the polarization products to Stokes parameters. More details of the calibration of the UWL data can be found in ~\cite{dlb+2019}.

We used the {\tt RMFIT} to determine the rotation measure (RM) for each observation, and subsequently corrected the profile based on its corresponding RM value. The {\tt PSRFLUX} was used to measure the flux density of each pulsar by cross-correlating the observed profile with a standard template. The uncertainty of the flux density was estimated by calculating the standard deviation of baseline fluctuations. For the polarization profile, we calculated fractional linear polarization ($f_{\rm L}$), net circular polarization ($f_{\rm C}$), and absolute circular polarization ($f_{\rm \lvert C \lvert}$) based on their average values, while estimating uncertainties using the root-mean-square (rms) noise from the baseline~\citep{dhm+2015}. To investigate the frequency-dependent evolution of emission states, we divided the entire bandwidth into {8 sub-bands of 416\,MHz width}, centered at frequencies of 912\,MHz, 1328\,MHz, 1744\,MHz, 2160\,MHz, 2576\,MHz, 2992\,MHz, 3408\,MHz and 3824\,MHz, respectively.

\begin{figure*}
\centering
\subfigure[PSR J0941$-$39]{
\includegraphics[width=0.92\columnwidth]{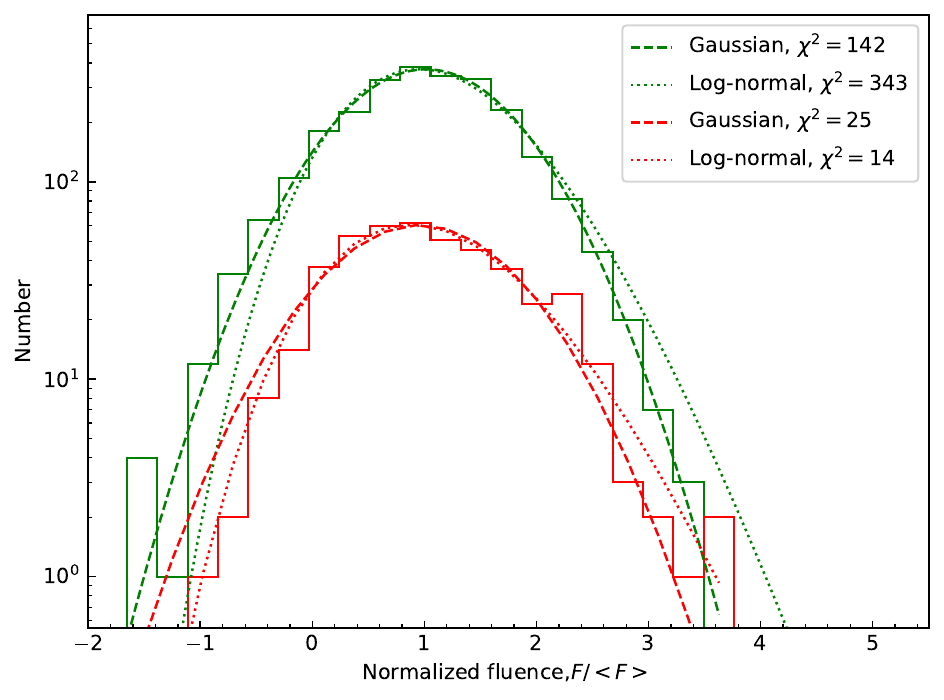}
 }
   \subfigure[PSR J1107$-$5907]{
\includegraphics[width=0.92\columnwidth]{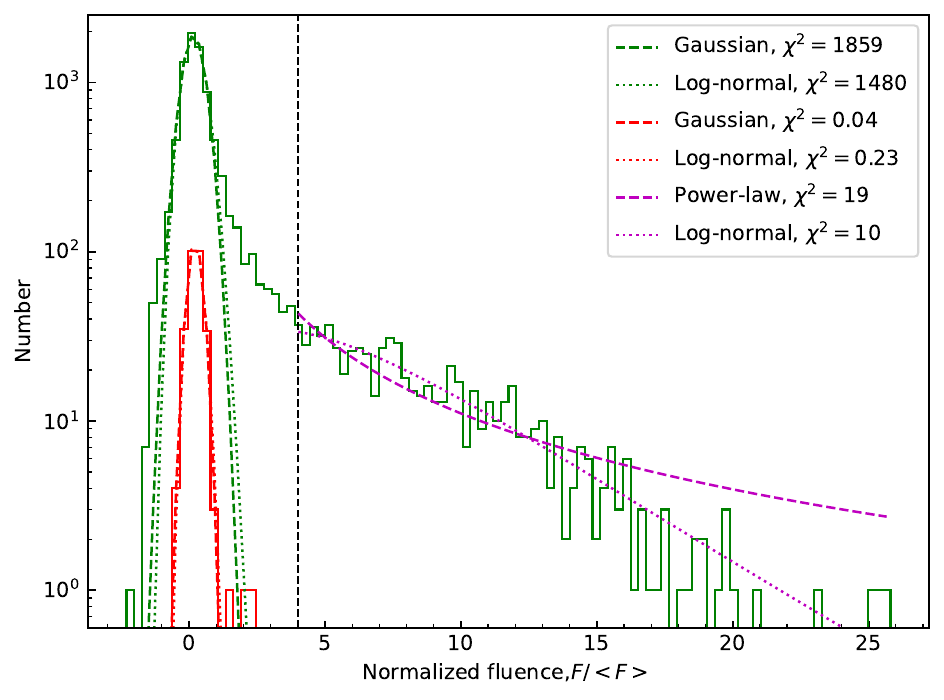}
  }
 \caption{Normalized fluence cumulative distributions for pulsar state pulses (green solid histogram) and RRAT state (red solid histogram) for PSR J0941$-$39 (panel (a)) and PSR J1107$-$5907 (panel (b)). The fluences are normalized by the mean on-pulse fluence $\langle F \rangle$. The dotted and dashed lines are for the fitting results. In which, the vertical black dashed line represents the $4 \langle F \rangle$.}
 \label{fig:1ehist}
 \end{figure*}

\begin{table*}
\renewcommand\arraystretch{1.3}
\centering
\caption{Best-fit parameters for the two state of PSR J0941$-$39 and PSR J1107$-$5907.}
\label{fit}
\begin{tabular}{cccccccccccccccc}
\hline                    
PSR  & State  & \multicolumn{4}{c}{Gaussian}  & \multicolumn{4}{c}{Log-normal}  &  \multicolumn{3}{c}{Power Law} &  \multicolumn{3}{c}{Log-normal} 
\\
\cline{3-5}\cline{7-9} \cline{11-12} \cline{14-16}
\\
& & $\mu$ & $\sigma$ & $\chi^2$ & & $\mu$ & $\sigma$ & $\chi^2$ & & $\beta$ & $\chi^2$ & & $\mu$ & $\sigma$ & $\chi^2$ 
\\
\hline
\multirow{2}{*}{J0941$-$39} & Pulsar & $1.01(1)$ & $0.73(1)$ & 142 & & $1.79(1)$ & $0.12(1)$ & 343 & & ... & ... &  & ... & ... & ...
\\
 & RRAT & $0.95(4)$ & $-0.79(4)$ & 25 & & $1.78(1)$ & $0.13(1)$ & 14 & & ... & ... &  & ... & ... & ...
 \\
 \multirow{2}{*}{J1107$-$5907} & Pulsar & $0.16(1)$ & $0.41(1)$ & 1859 & & $1.64(1)$ & $0.08(1)$ & 1480 & & $-1.49(8)$ & 19 &  & $2.29(5)$ & $0.44(4)$ & 10
\\
 & RRAT &$0.22(1)$ & $0.27(1)$ & 0.04 & & $1.65(1)$ & $0.05(1)$ & $0.23$ & & ... & ... &  & ... & ... & ...
\\
\hline
 \end{tabular}
\end{table*}

\section{RESULTS}\label {sec:res}

We used {\tt RMFIT} to determine the RM (RM$_{\rm obs}$) of PSR J0941$-$39 and PSR J1107$-$5907. The ionospheric RM contribution (RM$_{\rm ion}$) for each pulsar was measured using IONFR~\citep{ssh+2013,sshd+2013}. The ISM RM contribution $\rm RM_{\rm ISM}=RM_{\rm obs}-RM_{\rm ion}$. Our results are shown in Table~\ref{obs2}. For J0941$-$39, we obtained the RM measurement for the first time. For PSR J1107$-$5907, our determined value of $\rm RM_{\rm ISM}$ is $22.6 (1) \, {\rm rad\,m^{-2}}$, which agrees with the reported value of $23 (3) \, {\rm rad\,m^{-2}}$ by~\cite{yws+2014}, but slightly smaller than the reported value of $25.9 (3) \, {\rm rad\,m^{-2}}$ by~\cite{mtb+2018}. The profile of each pulsar was corrected using our measured values of RM.

\begin{figure*}
 \centering
 \subfigure[PSR J0941$-$39]{
 \begin{minipage}[t]{\textwidth}
 \centering
\includegraphics[width=0.18\columnwidth]{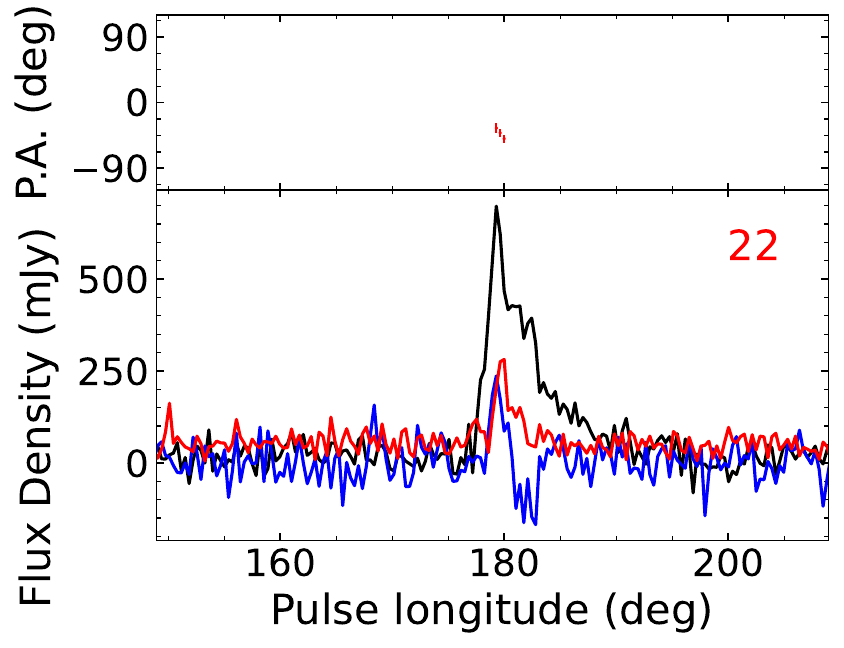}
\includegraphics[width=0.18\columnwidth]{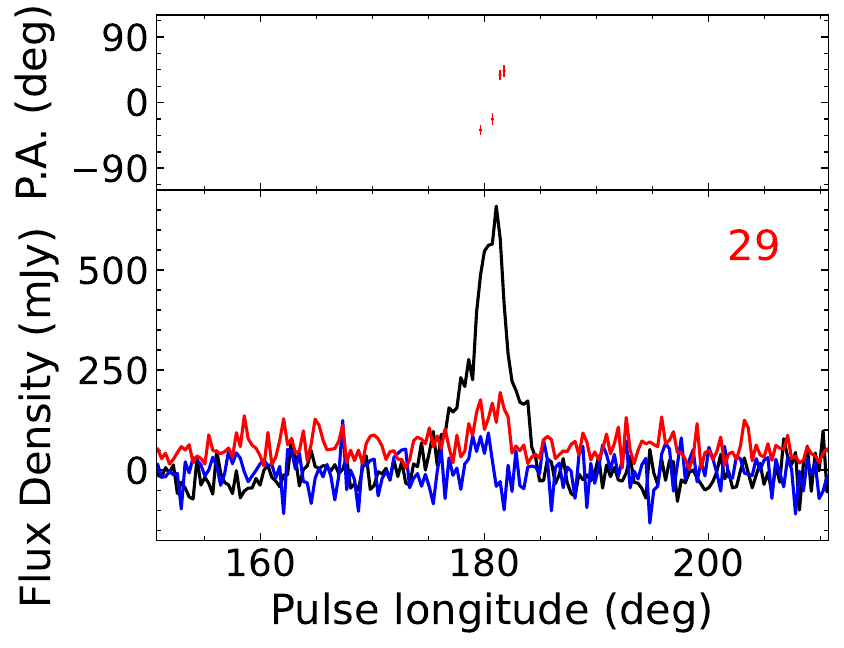}
\includegraphics[width=0.18\columnwidth]{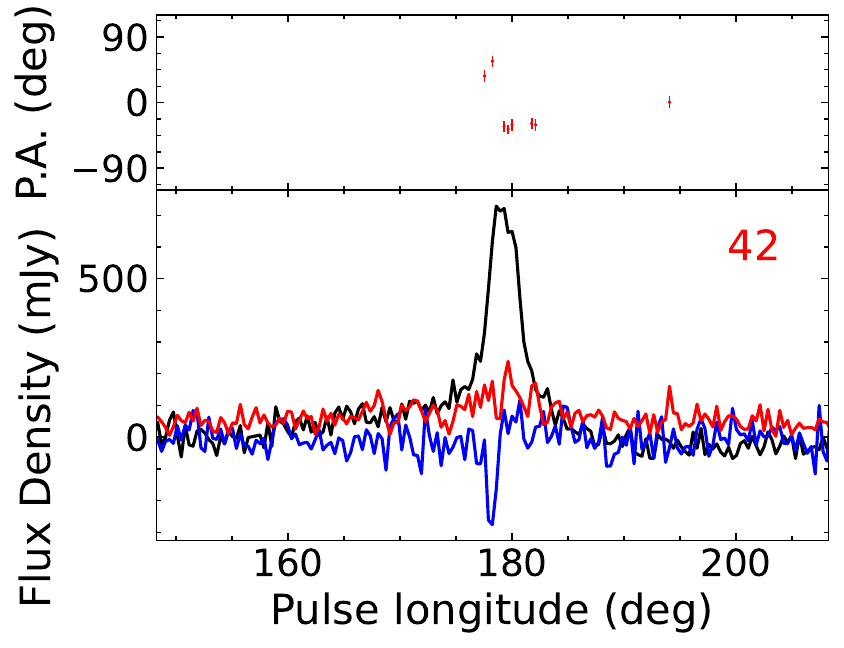}
\includegraphics[width=0.18\columnwidth]{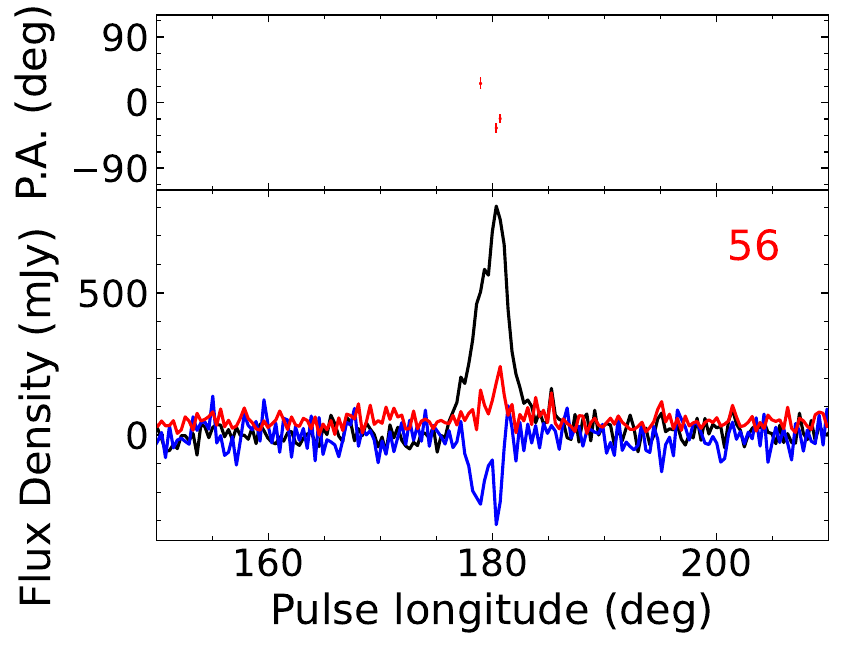}
\includegraphics[width=0.18\columnwidth]{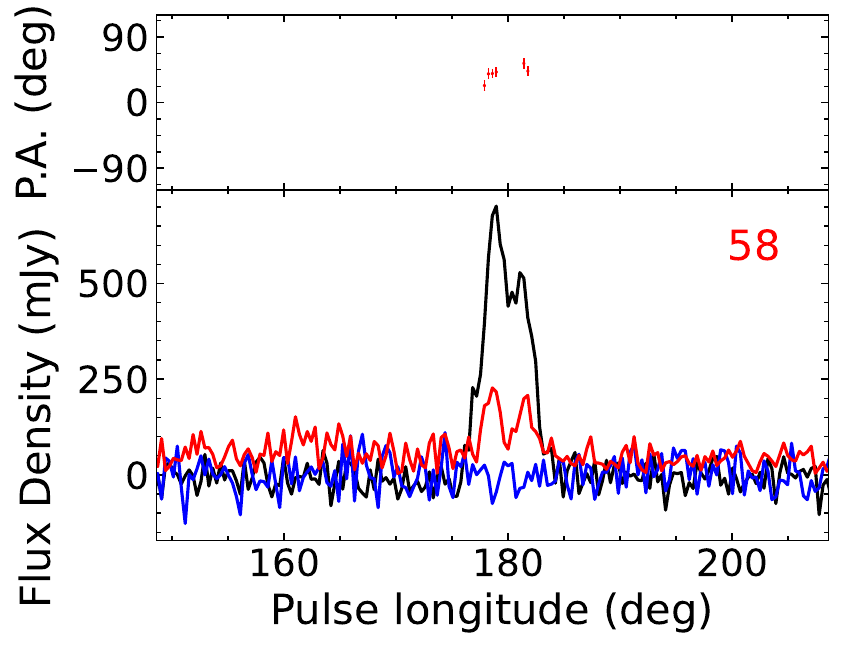}
\includegraphics[width=0.18\columnwidth]{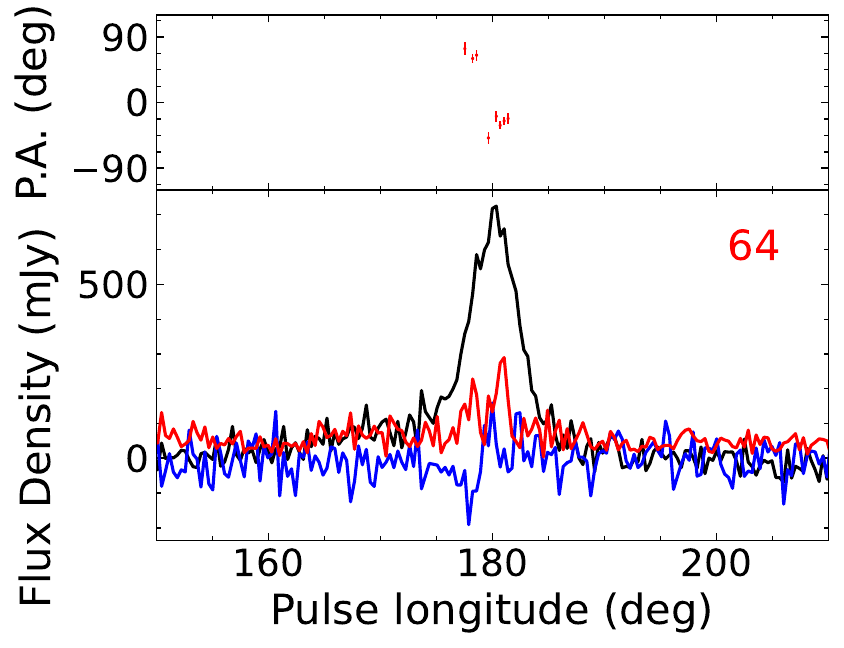}
\includegraphics[width=0.18\columnwidth]{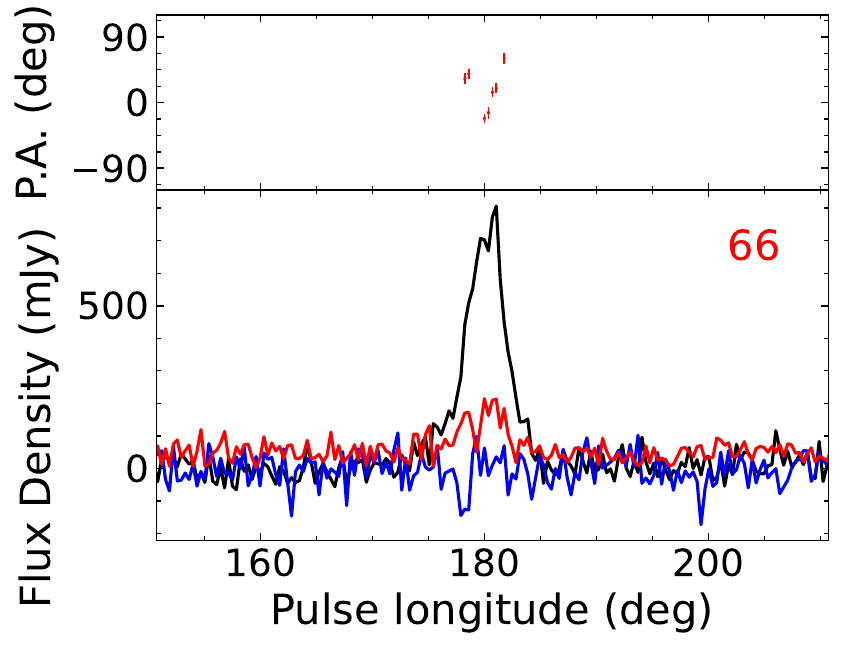}
\includegraphics[width=0.18\columnwidth]{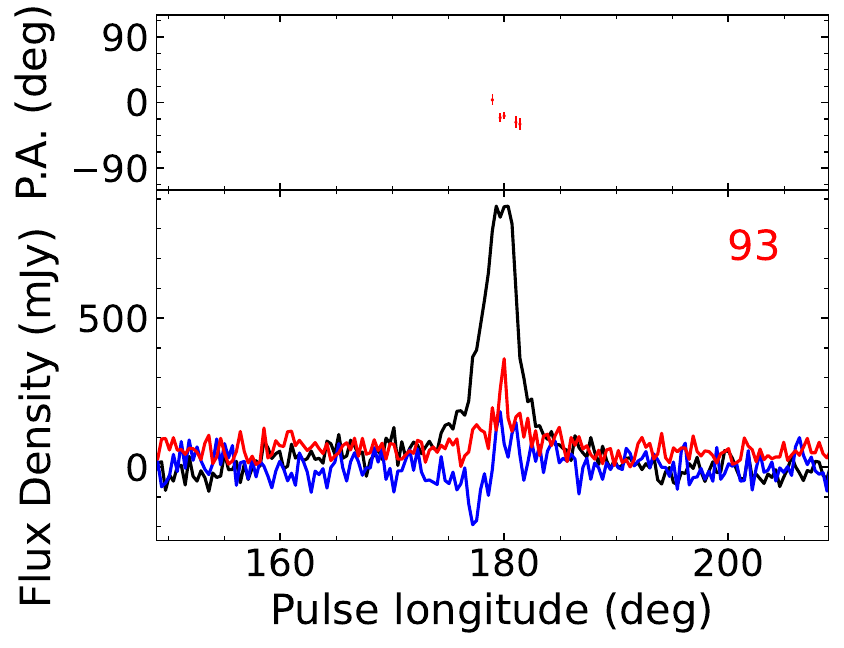}
\includegraphics[width=0.18\columnwidth]{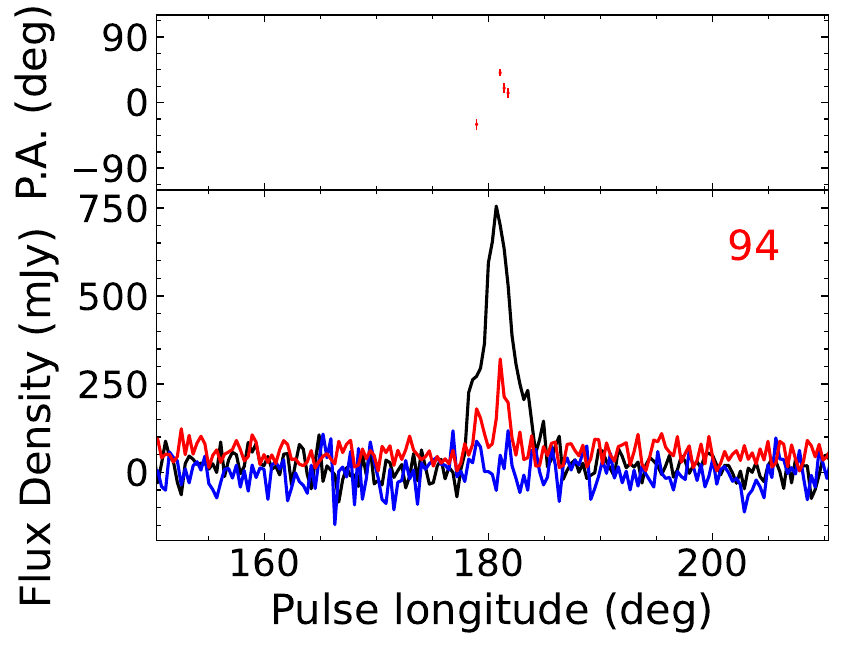}
\includegraphics[width=0.18\columnwidth]{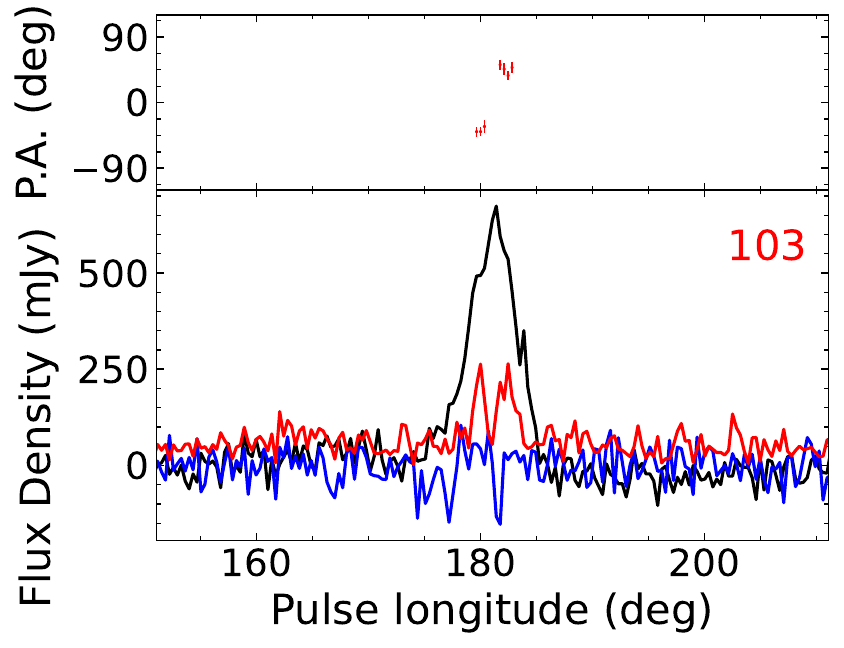}
\includegraphics[width=0.18\columnwidth]{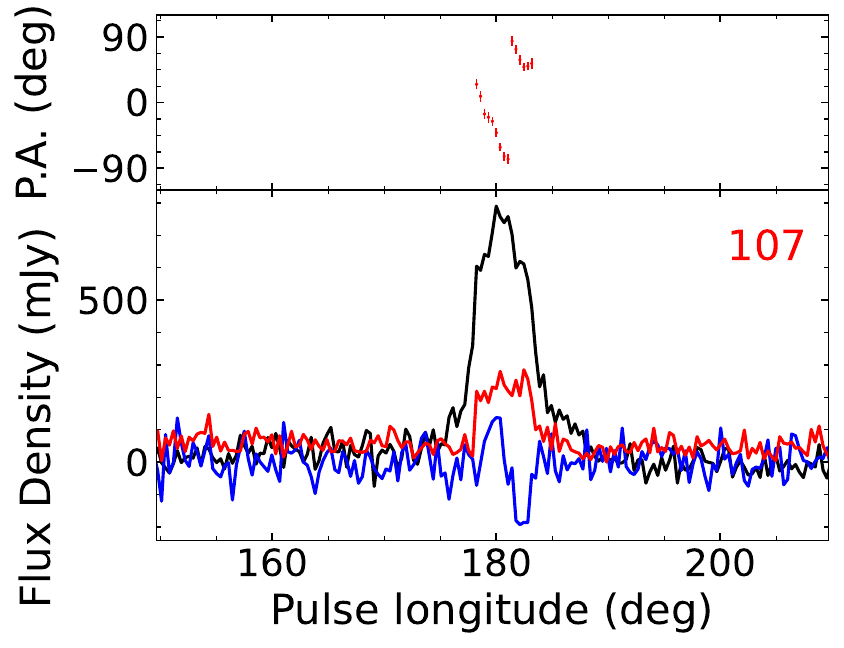}
\includegraphics[width=0.18\columnwidth]{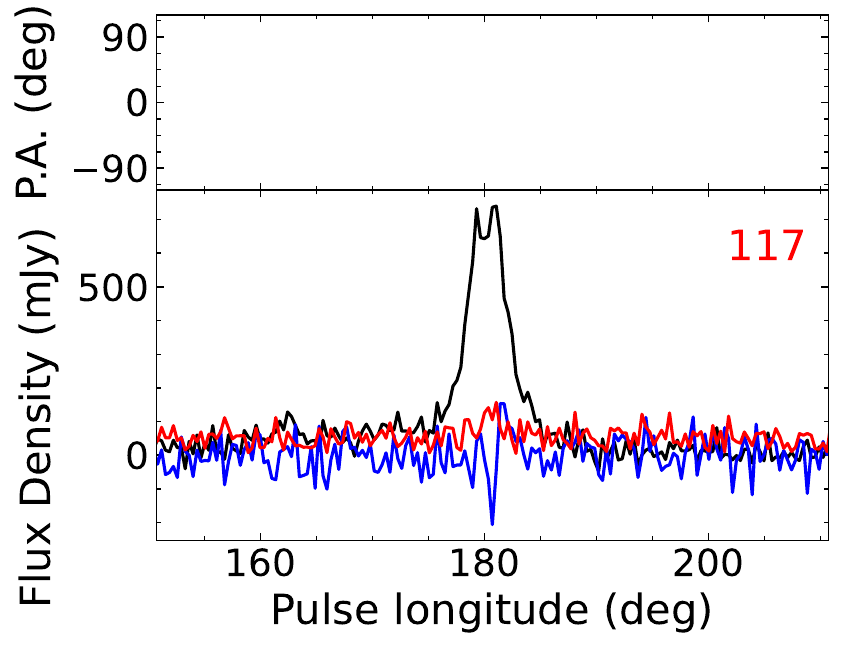}
\includegraphics[width=0.18\columnwidth]{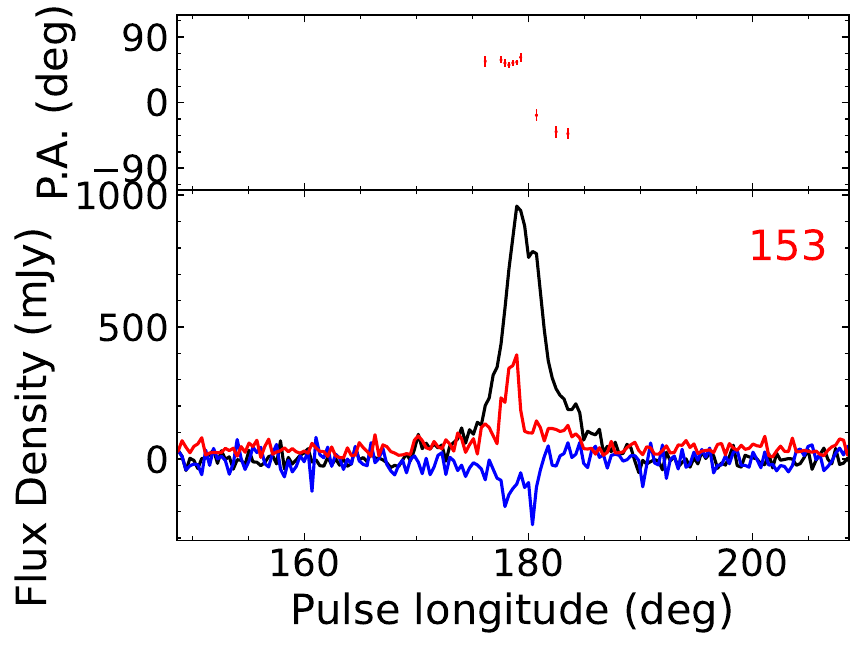}
\includegraphics[width=0.18\columnwidth]{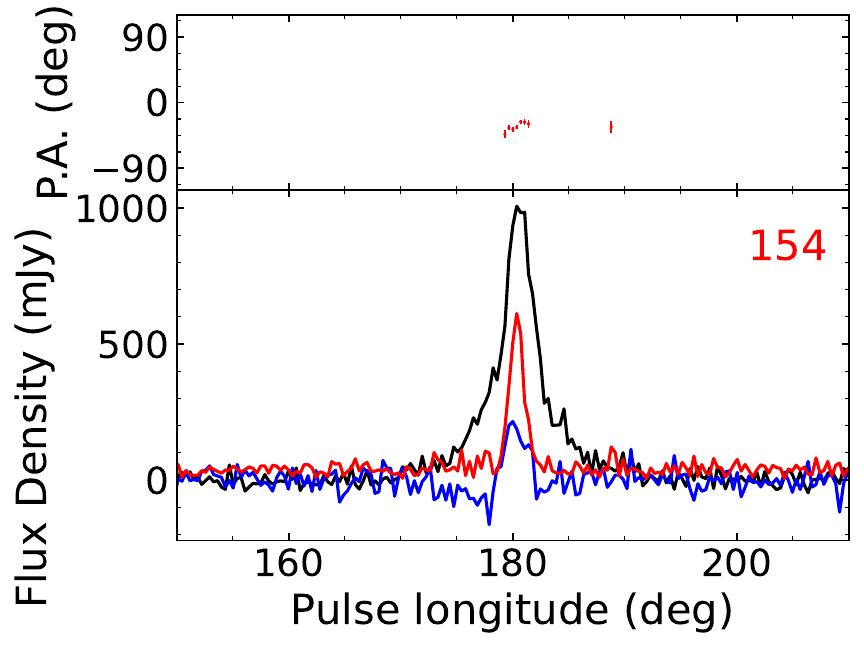}
\includegraphics[width=0.18\columnwidth]{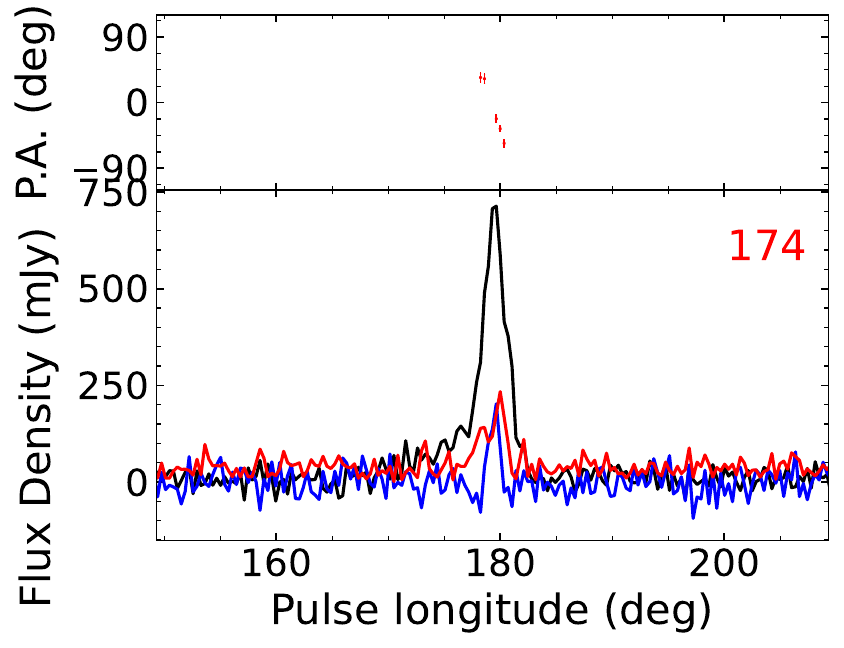}
\includegraphics[width=0.18\columnwidth]{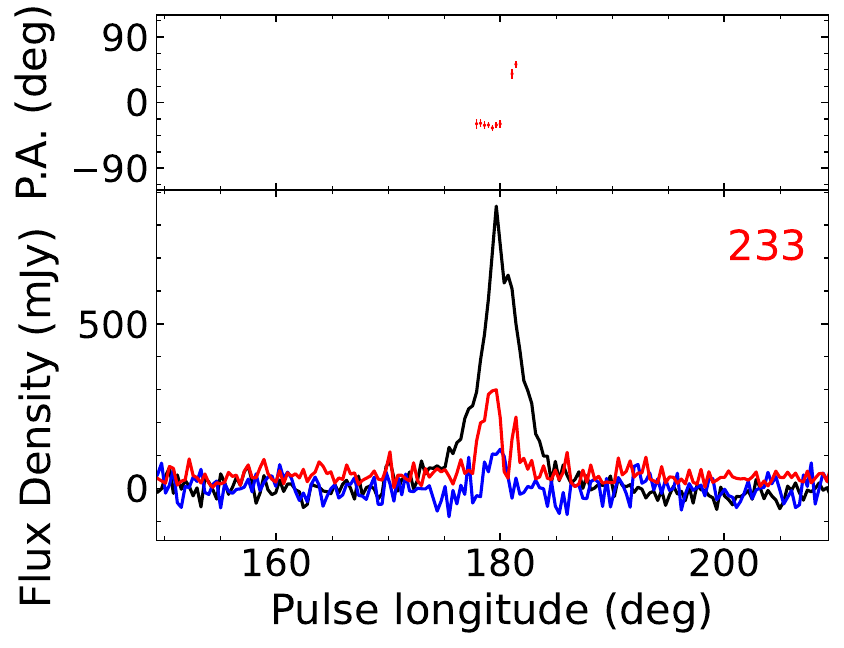}
\includegraphics[width=0.18\columnwidth]{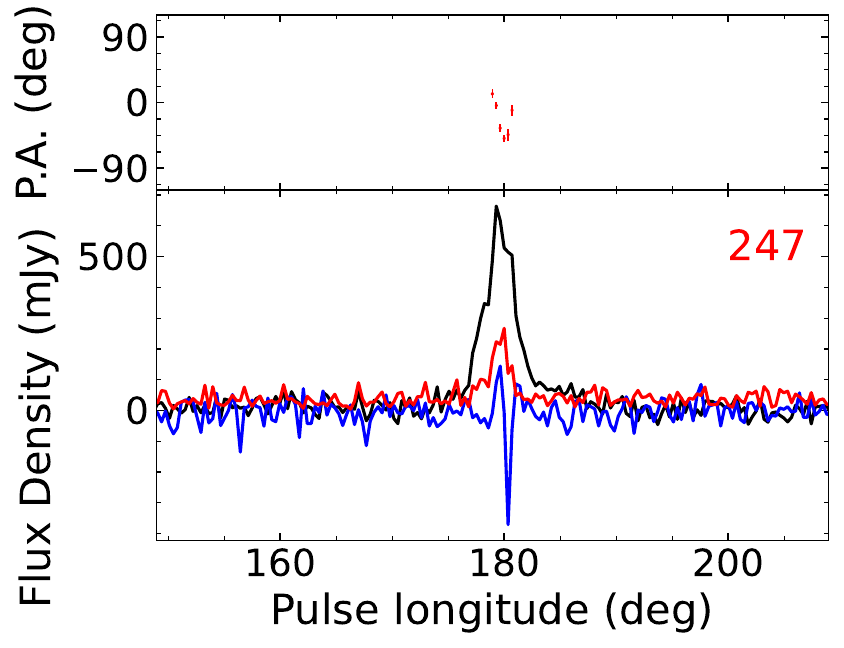}
\includegraphics[width=0.18\columnwidth]{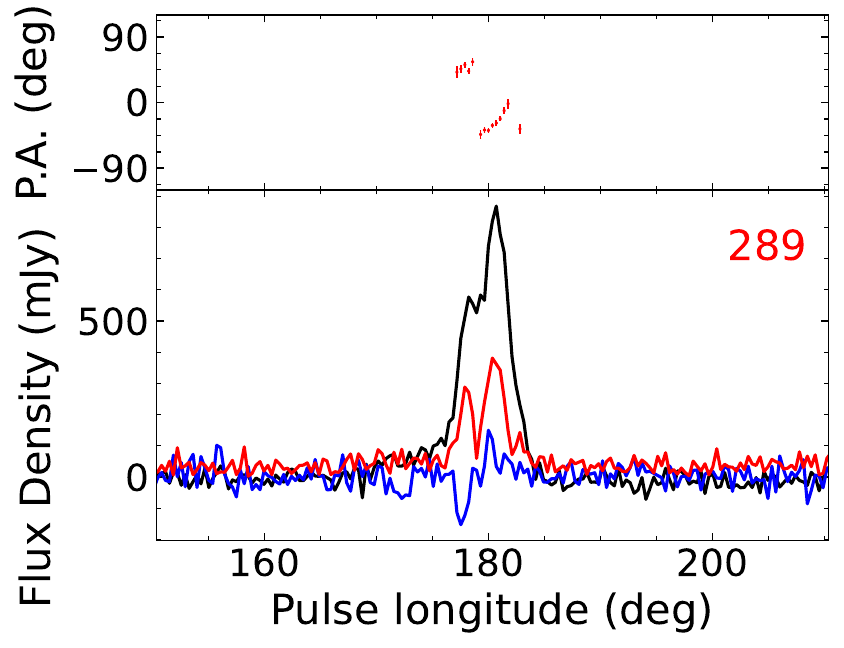}
\includegraphics[width=0.18\columnwidth]{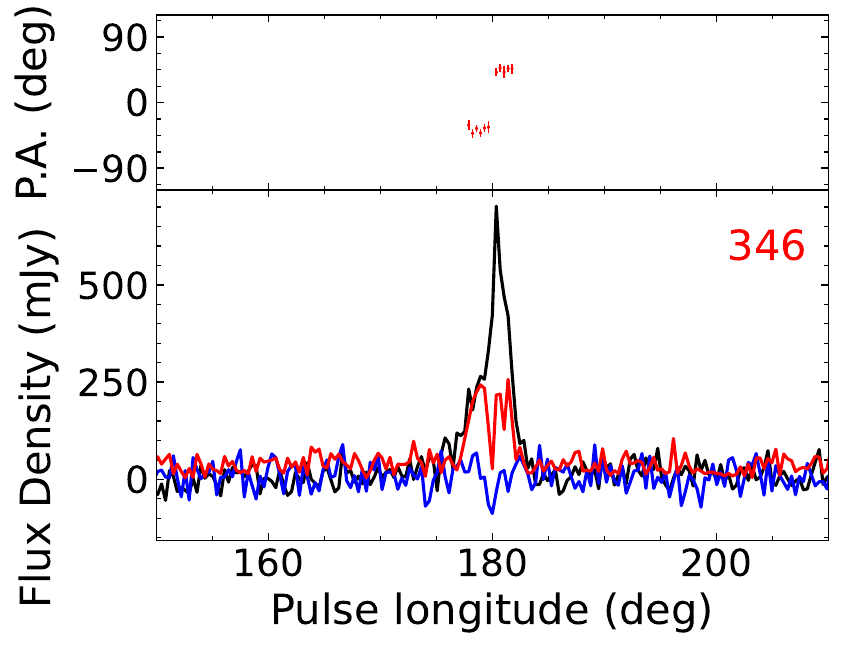}
\includegraphics[width=0.18\columnwidth]{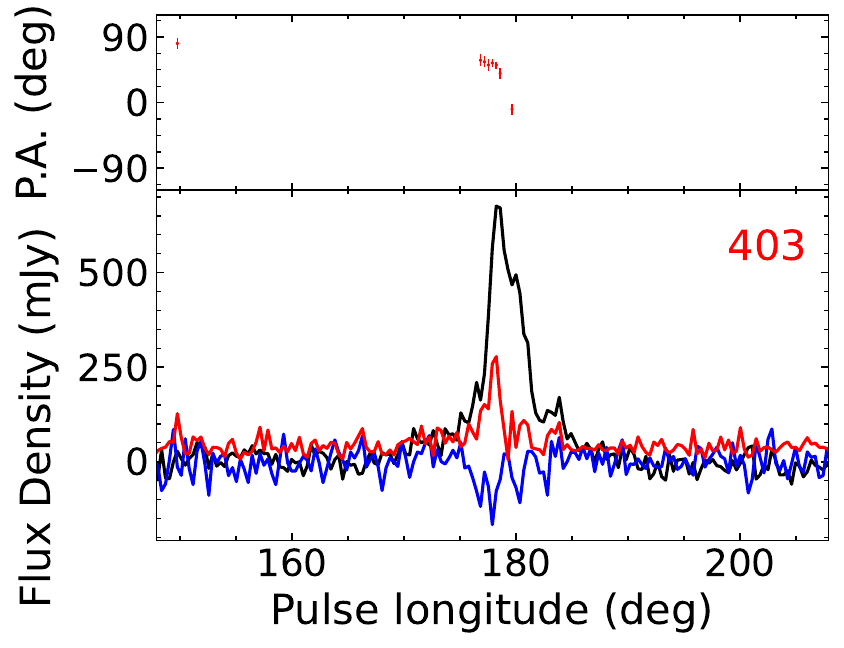}
 \end{minipage}
  }  
 \subfigure[PSR J1107$-$5907]{
 \begin{minipage}[t]{\textwidth}
  \centering
 \includegraphics[width=0.18\columnwidth]{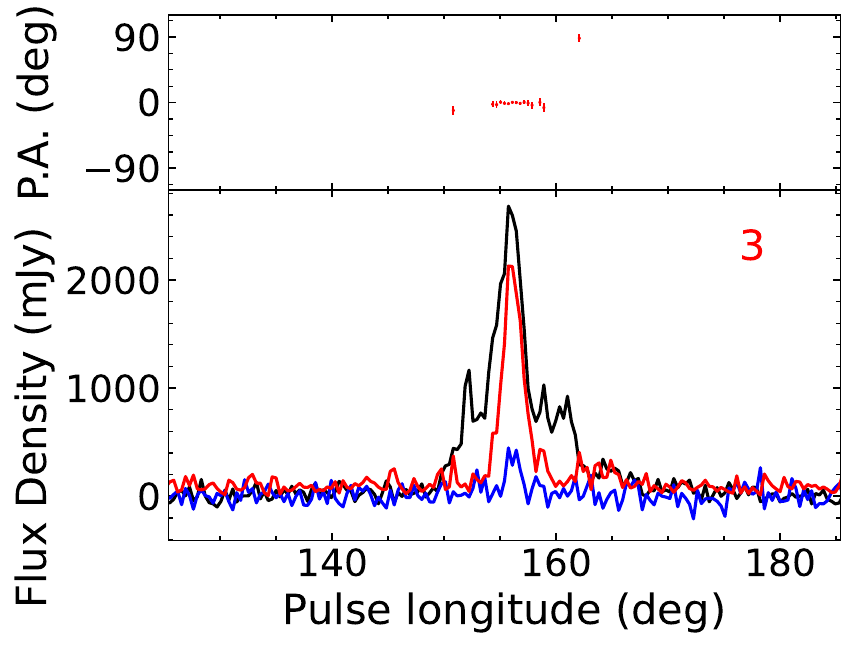}
\includegraphics[width=0.18\columnwidth]{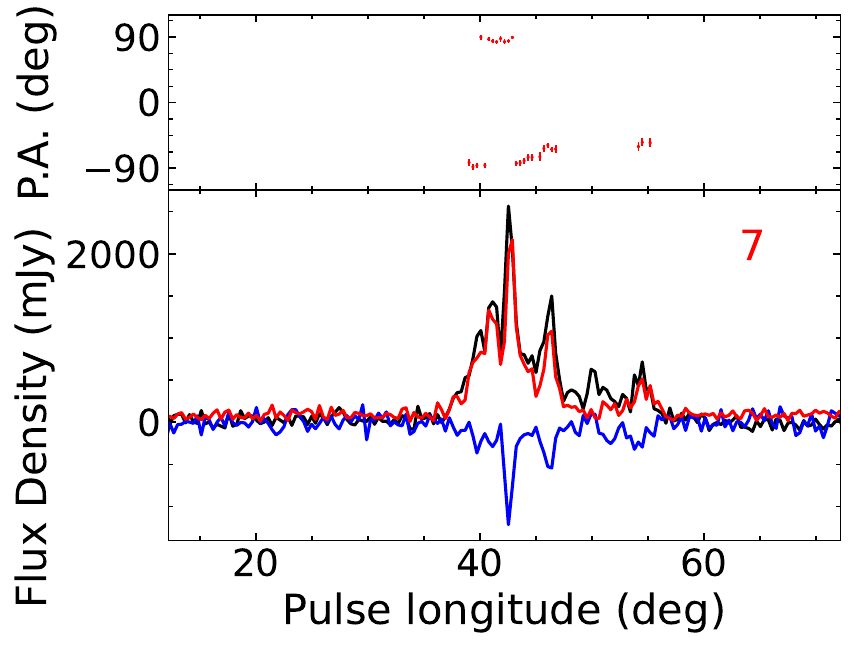}
\includegraphics[width=0.18\columnwidth]{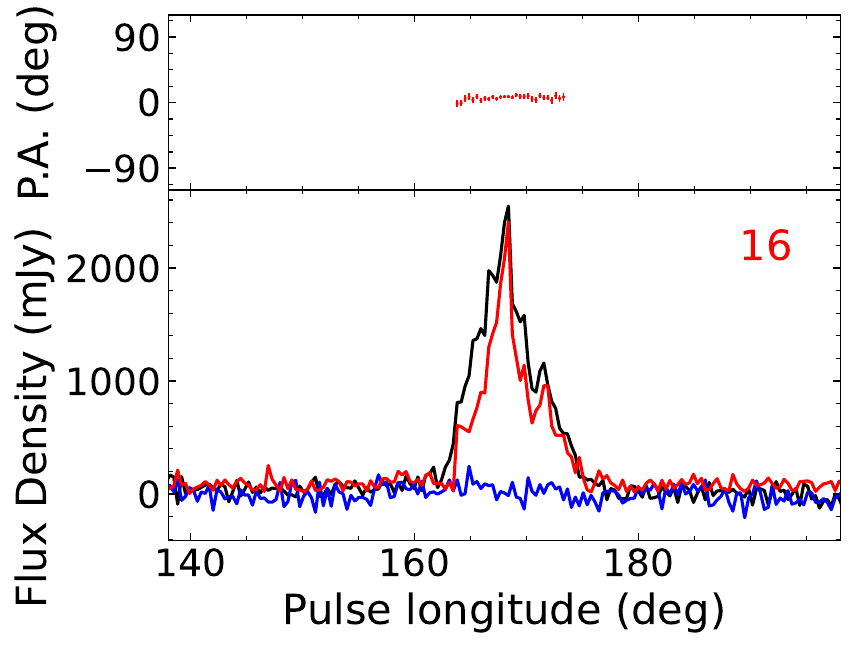}
\includegraphics[width=0.18\columnwidth]{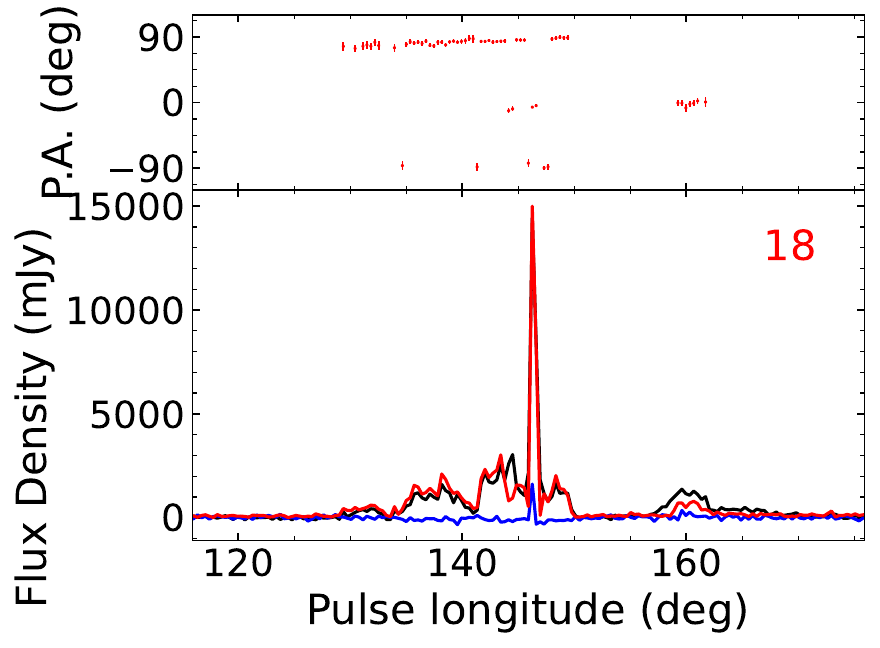}
\includegraphics[width=0.18\columnwidth]{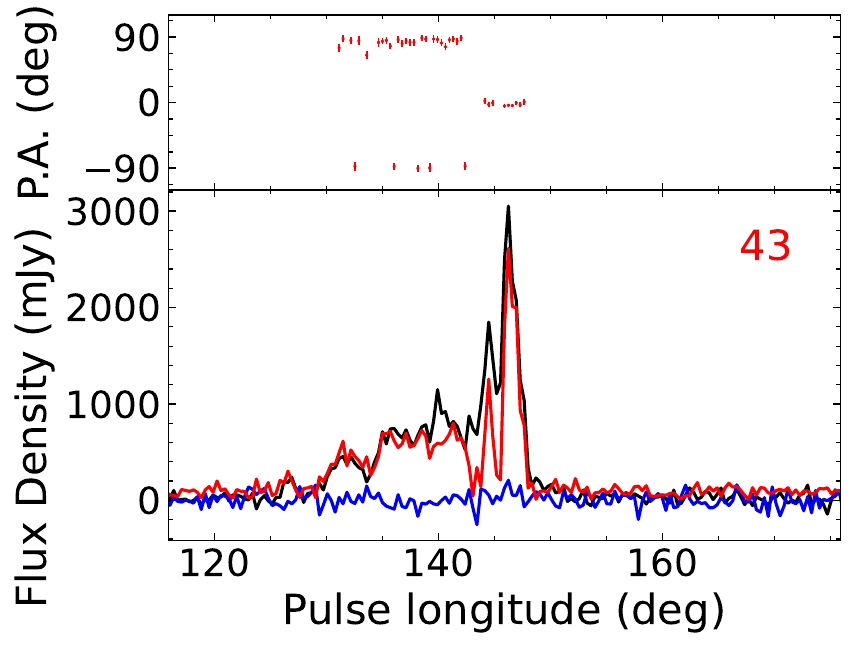}
\includegraphics[width=0.18\columnwidth]{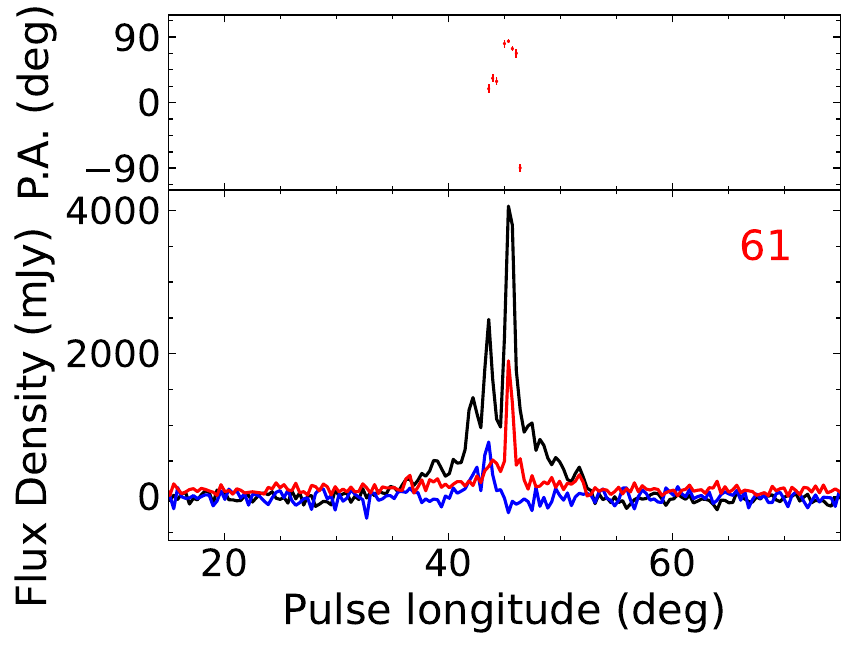}
\includegraphics[width=0.18\columnwidth]{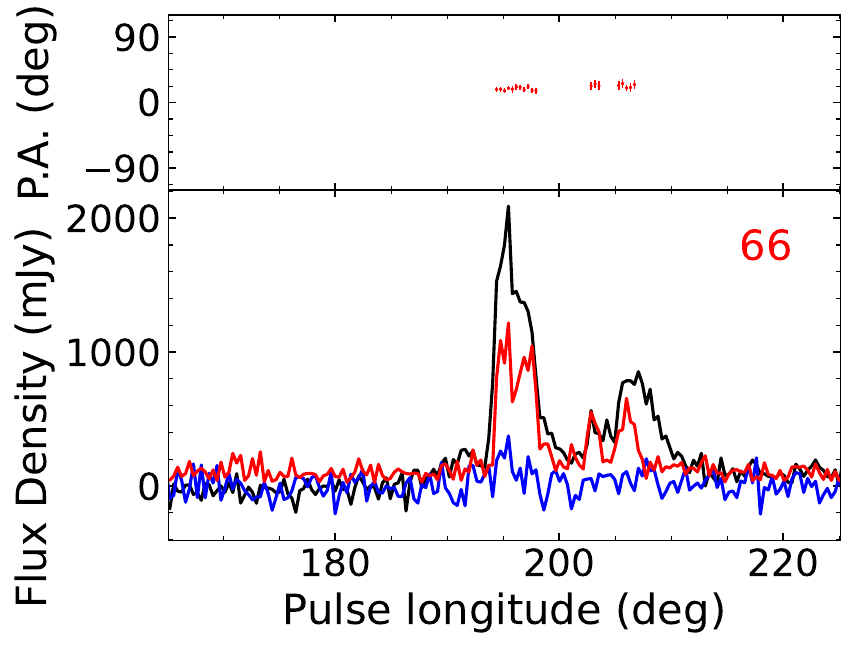}
\includegraphics[width=0.18\columnwidth]{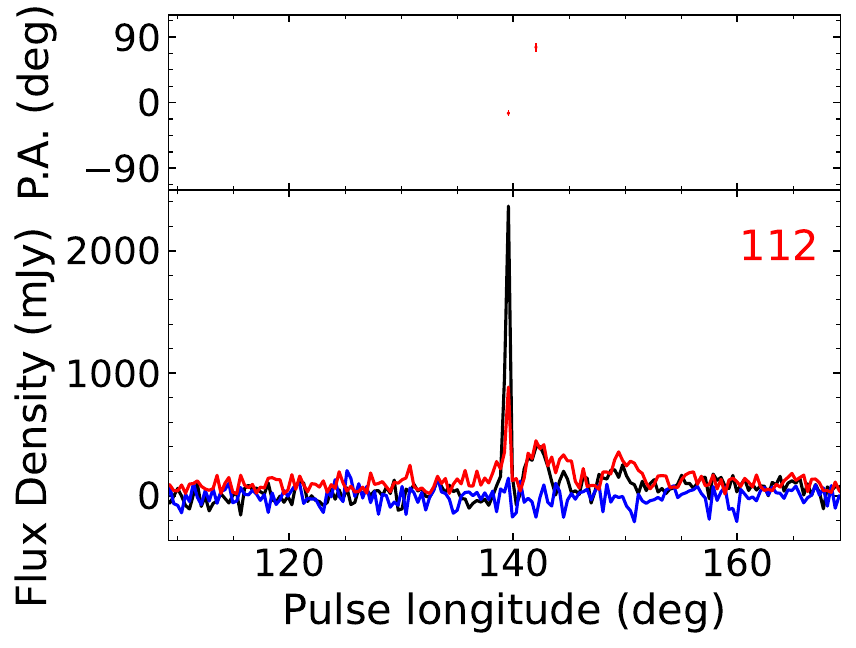}
\includegraphics[width=0.18\columnwidth]{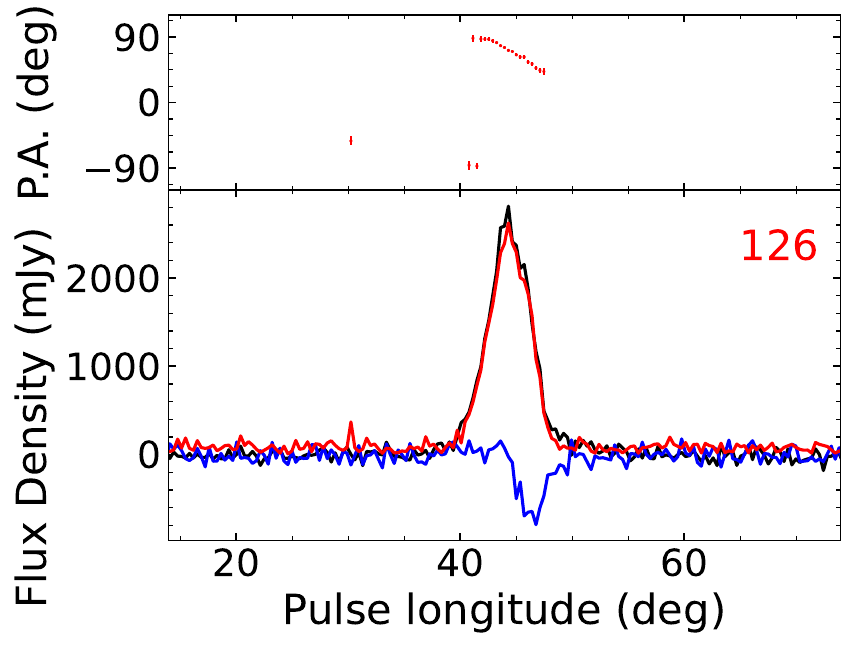}
\includegraphics[width=0.18\columnwidth]{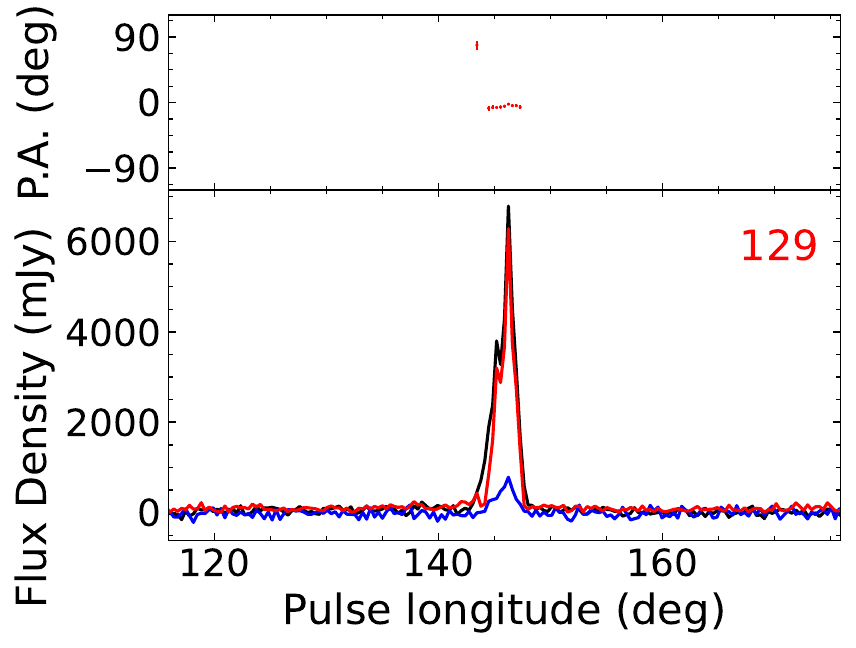}
\includegraphics[width=0.18\columnwidth]{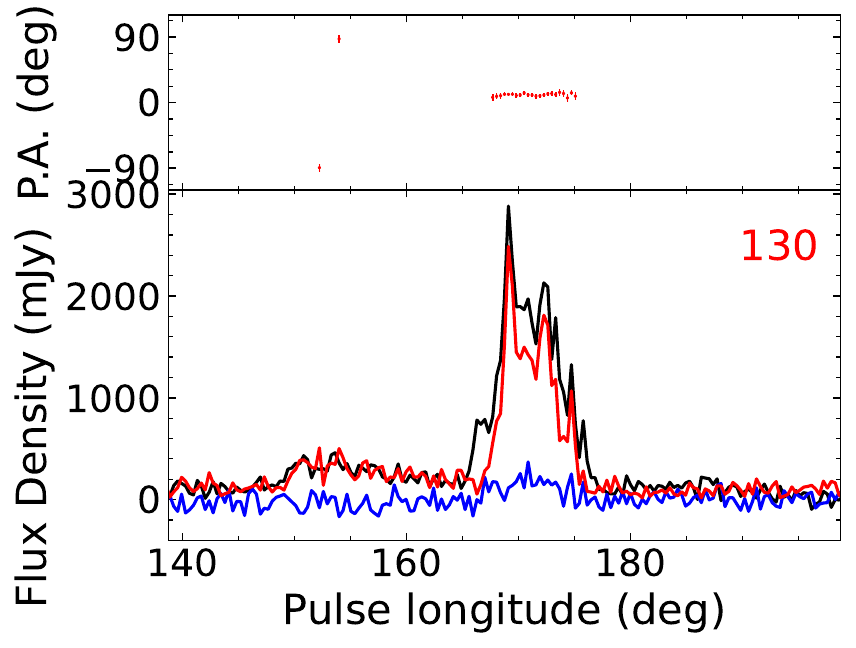}
\includegraphics[width=0.18\columnwidth]{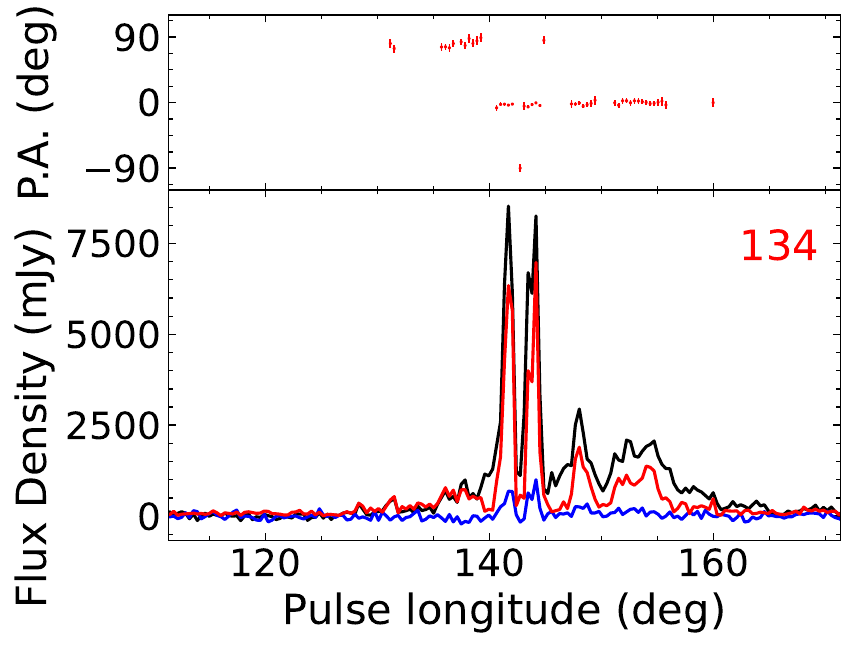}
\includegraphics[width=0.18\columnwidth]{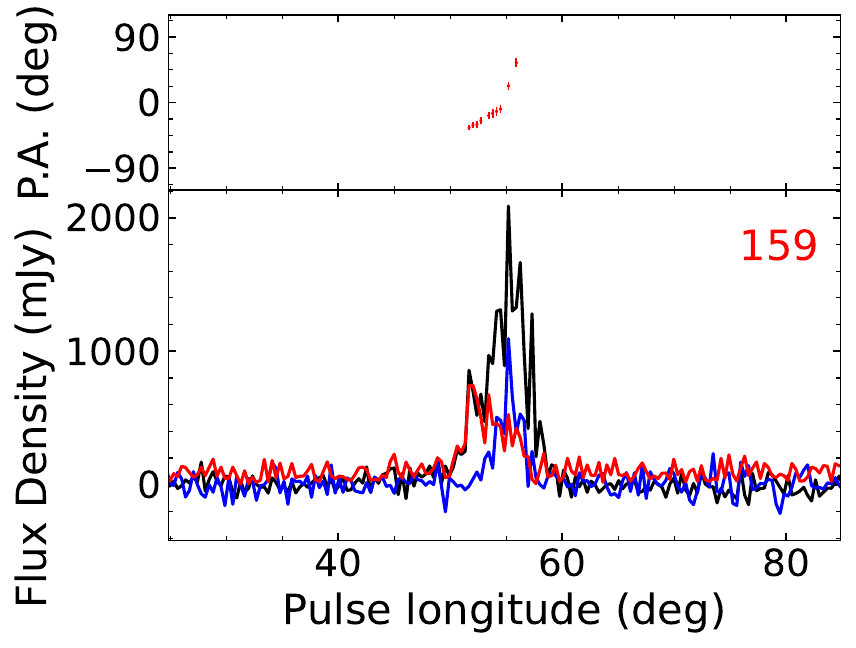}
\includegraphics[width=0.18\columnwidth]{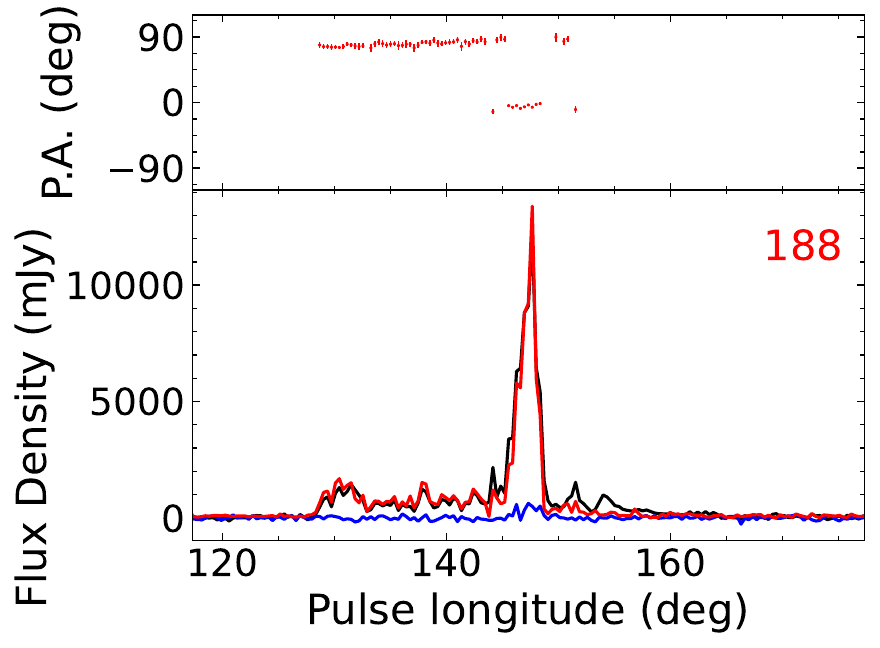}
\includegraphics[width=0.18\columnwidth]{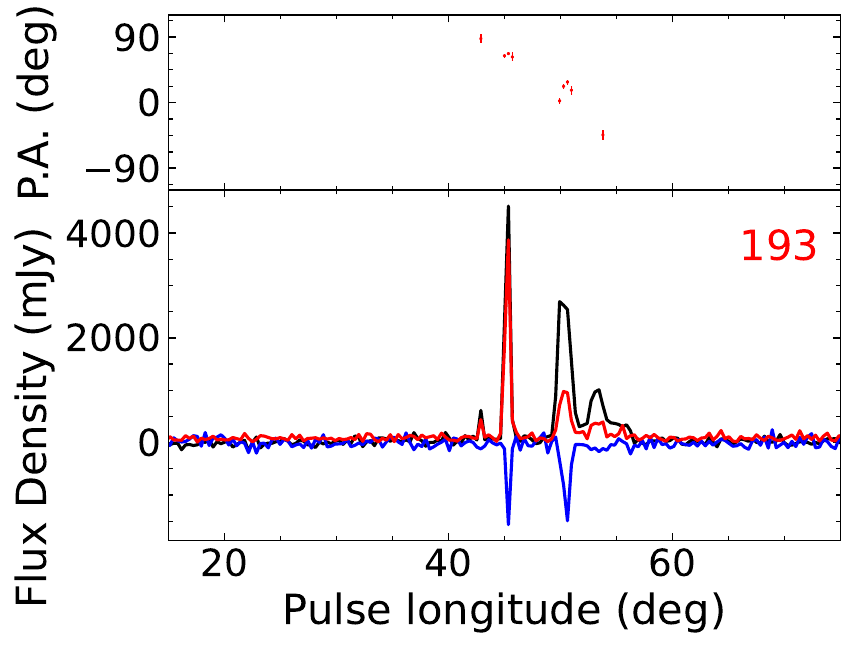}
\includegraphics[width=0.18\columnwidth]{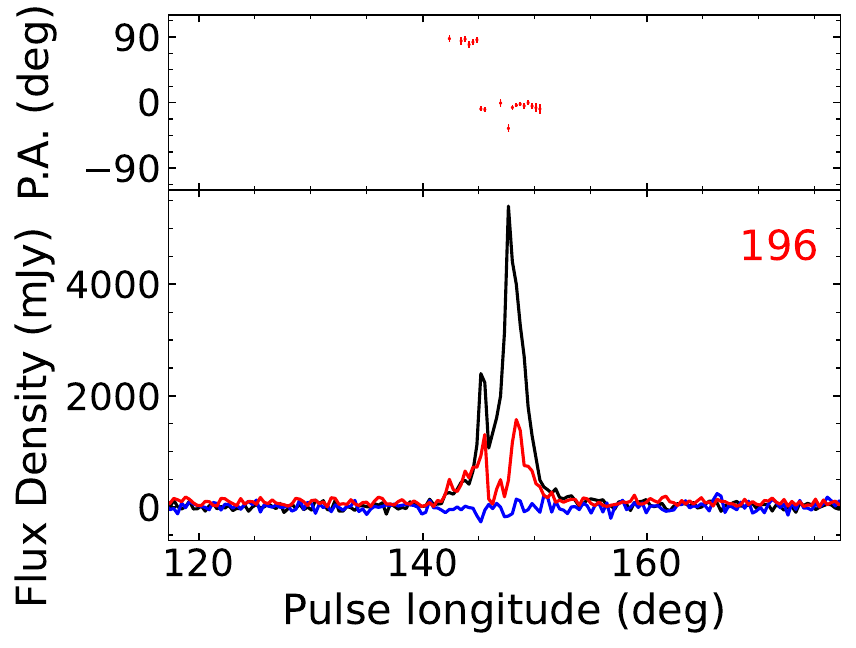}
\includegraphics[width=0.18\columnwidth]{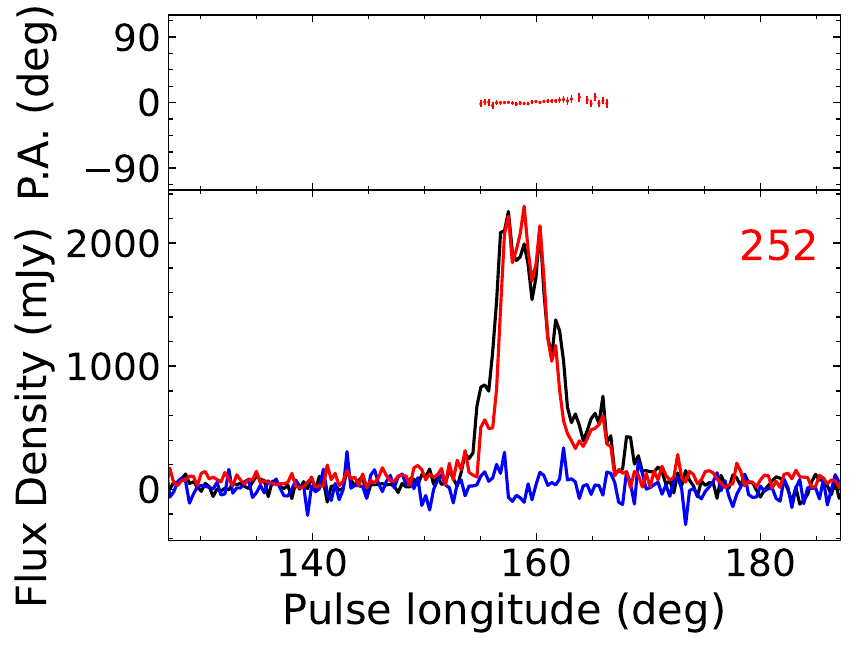}
\includegraphics[width=0.18\columnwidth]{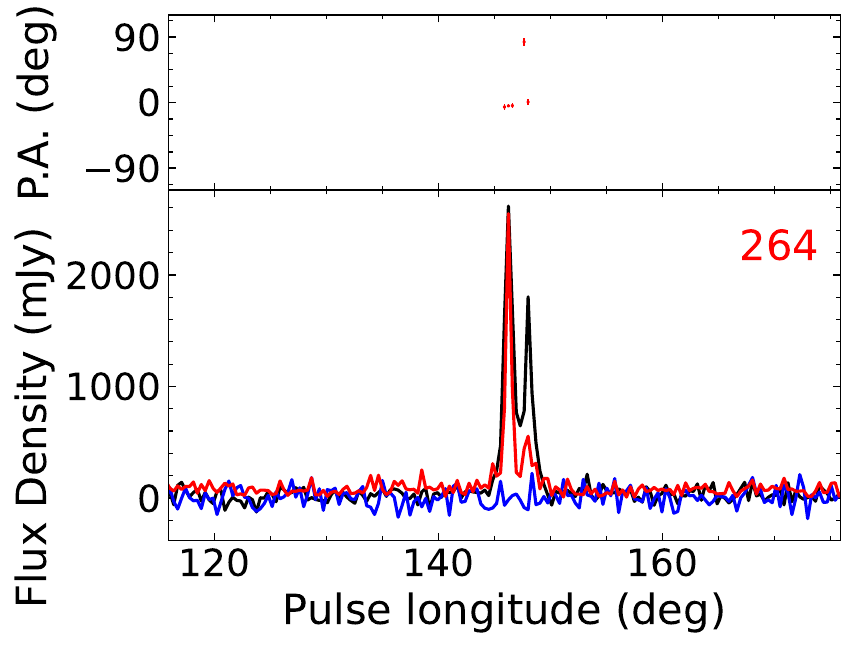}
\includegraphics[width=0.18\columnwidth]{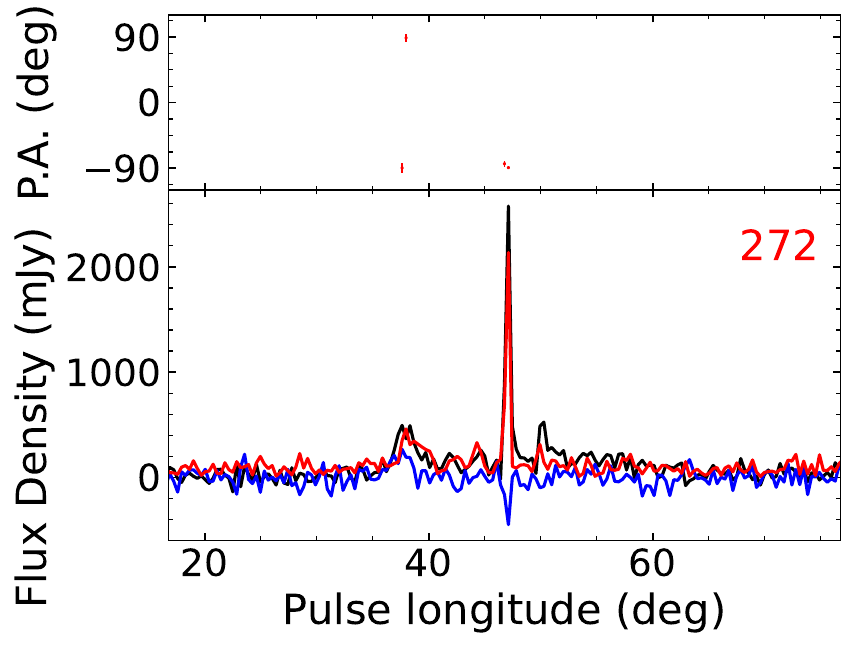}
\includegraphics[width=0.18\columnwidth]{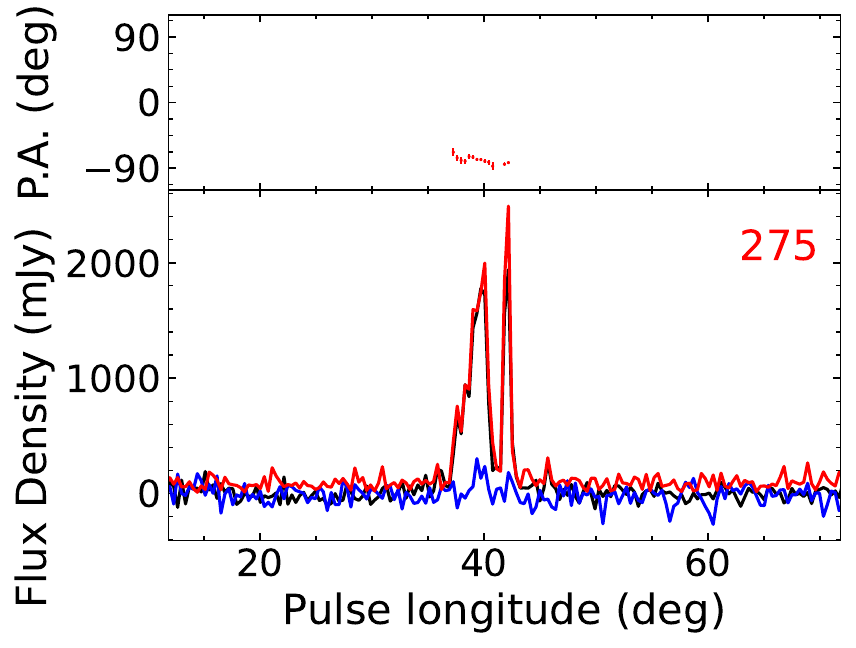}
 \end{minipage}
  } 
 \caption{The polarization profiles for the brightest 20 single pulses in RRAT states for PSR J0941$-$39 (panel (a)) and PSR J1107$-$5907 (panel (b)). The labels are the same as in Figure~\ref{fig:1pa}.}
 \label{fig:1singlepulses}
\end{figure*}

\subsection{Two emission states}

\subsubsection{PSR J0941$-$39}

We carried out three observations of PSR J0941$-$39, during which the pulsar exhibited the RRAT state on January 19, 2022 and November 14, 2022, while displaying the pulsar state on June 14, 2022. Panel (a) of Figure~\ref{fig:1state} presents some single pulses of PSR J0941$-$39
during both states. Sporadic single pulses were detected during the RRAT state.

To construct the profile of RRAT state, we selected single pulses with a peak signal-to-noise ratio $S/N_{\rm p}>7$. Here, $S/N_{\rm p}$ is calculated as the ratio of the maximum intensity within the on-pulse region to the rms in the off-pulse region. It should be noted that some authors commonly use threshold values of S/N of 6 (e.g., \citet{mtb+2019}) or 5 (e.g.,~\citet{xww+2022,cbm+2017}) for classifying single pulses in RRATs. During our observations, we noticed that certain single pulses exhibited narrow characteristics during RRAT states. Therefore, we adopted a criterion of $S/N_{\rm p}>7$ for classification purposes while setting a higher threshold value of 7 to mitigate  potential single pulse candidates induced by RFI. A total of 440 single pulses with $S/N_{\rm p}>7$ were detected across two observations during RRAT states. Note that we carefully examined frequency-phase plots of these pulses to eliminate any possible RFI contamination.

The frequency-averaged polarization profiles of the pulsar state and RRAT state are presented in panel (a) of Figure~\ref{fig:1pa}. The average pulse profile of the pulsar state exhibits three components, with a notably stronger third component. The measured values for $f_{\rm L}$, $f_{\rm C}$ and $f_{\rm \lvert C \lvert}$ are $38.7 (1)\%$, $0.5(5)\%$ and $8.6(5)\%$, respectively. A reversal in circularly polarized intensity is observed at the phase of 177\,deg. The average pulse profile of the RRAT state also displays three components with a stronger third component. We measure $f_{\rm L}$, $f_{\rm C}$, and $f_{\rm \lvert C \lvert}$ as $17.1(1)\%$, $-0.4(4)\%$, and $10.9(4)\%$ respectively for RRAT state. Similarly to the pulsar state, there is also a reversal in circularly polarized intensity at the phase of 177\,deg. The polarization angle (PA) swings within their respective average profiles in both pulsar and RRAT states exhibit similar S-shaped variations. A jump by 90\,deg is detected at the pulse phase of 180\, deg for both states, which suggests the existence of orthogonal polarization modes (OPMs) phenomenon.

The pulse fluence (or energy) distributions ~\citep{mtb+2018,jcb+2023} for the pulsar state and RRAT state are shown in panel (a) of Figure~\ref{fig:1ehist}. The pulse fluence is computed by summing the intensities within the on-pulse window of the average pulse profile after subtracting the baseline noise. Gaussian and log-normal functions were used to fit the fluence distribution of each state. The fitting parameters for both states can be found in Table~\ref{fit}. By comparing the $\chi^2$ values, it was observed that the pulse fluence for the pulsar state follows a Gaussian distribution (represented by the green dashed line in panel (a) of Figure~\ref{fig:1ehist}), while that for RRAT state follows a log-normal distribution (depicted by the red dotted line in panel (a) of Figure~\ref{fig:1ehist}).

In Figure~\ref{fig:1singlepulses}, we present a sample of polarization profiles showcasing the 20 brightest single pulses observed during the RRAT state. These single pulses exhibit narrow widths, with full width at half maximum (W50) ranging from 3.09\,ms to 8.87\,ms. The profiles generally exhibit one or two peaks. Both $f_{\rm L}$ and $f_{\rm C}$ demonstrate significant variations within the ranges of about $7\%$ to $36\%$ and $-22\%$ to $6\%$, respectively. Different single pulses exhibit varying swings in PA. The peak intensity of the brightest single pulse observed during the RRAT state is approximately 6.5 times higher than that of the average pulsar profile in pulsar state.

 \begin{figure*}
 \centering
  \subfigure[PSR J0941$-$39]{
 \begin{minipage}[t]{0.92\textwidth}
  \centering
   \includegraphics[width=0.2\columnwidth]{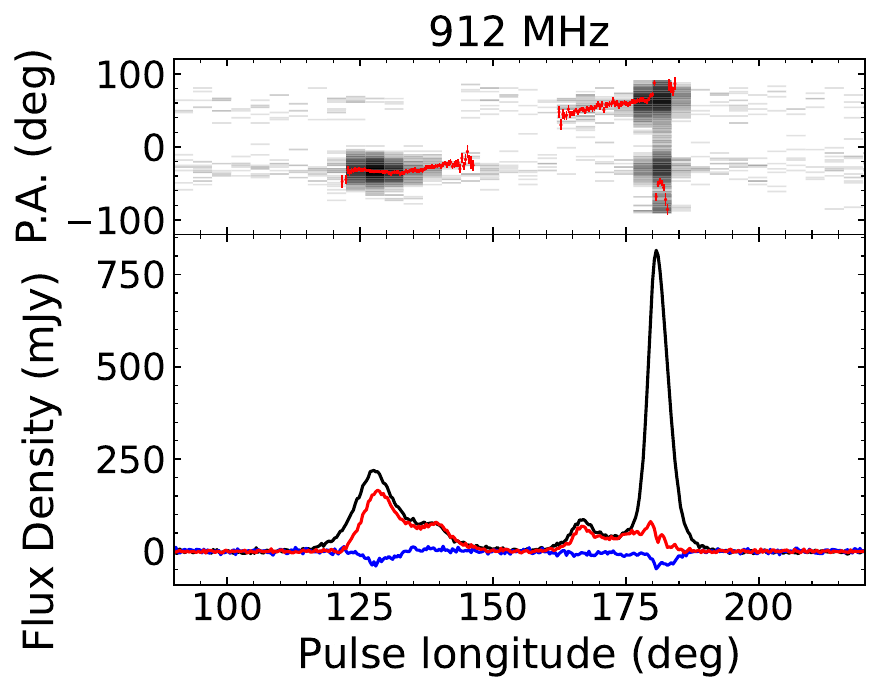}
   \includegraphics[width=0.2\columnwidth]{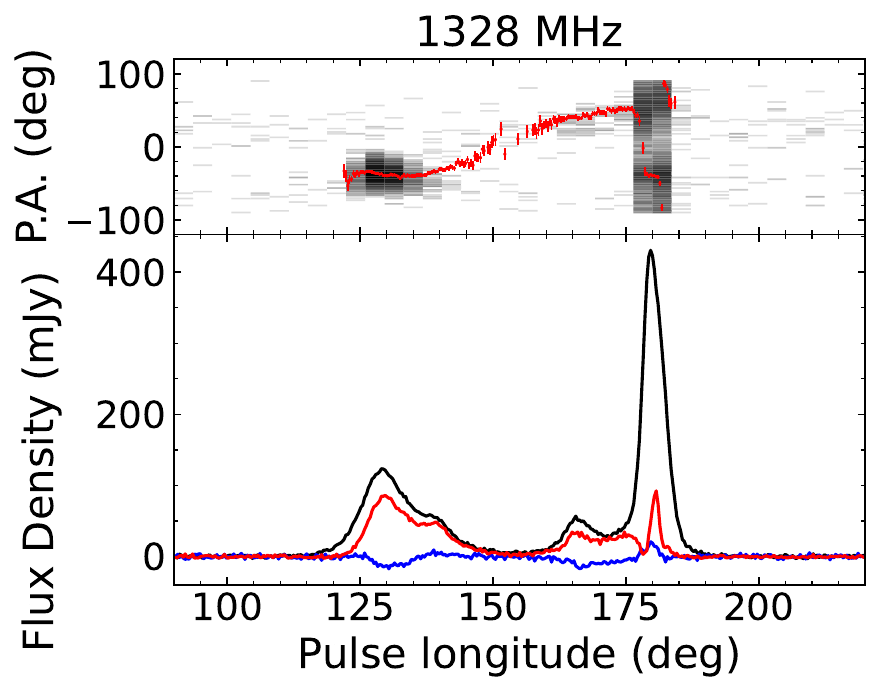}
   \includegraphics[width=0.2\columnwidth]{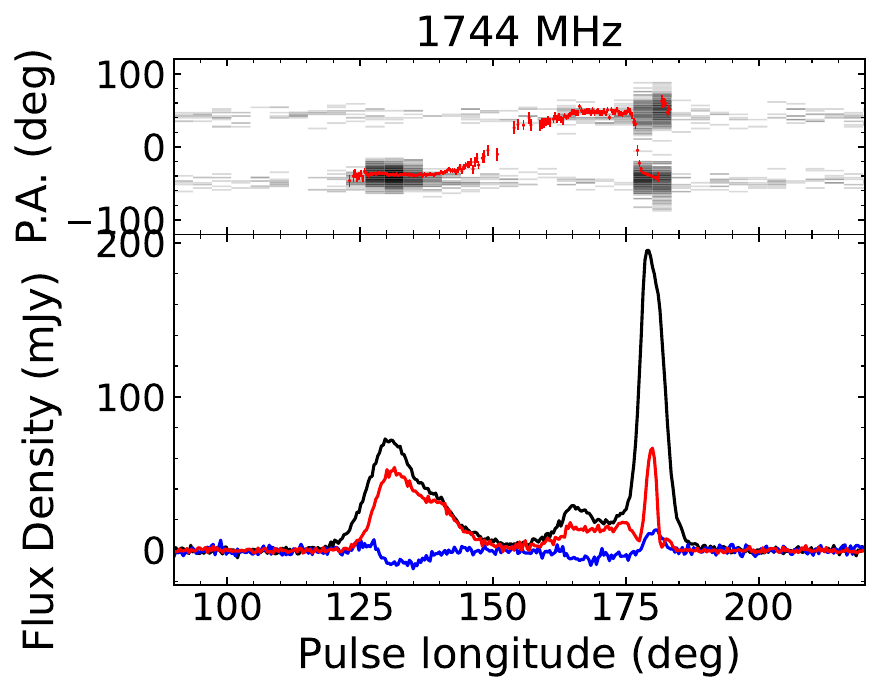}
   \includegraphics[width=0.2\columnwidth]{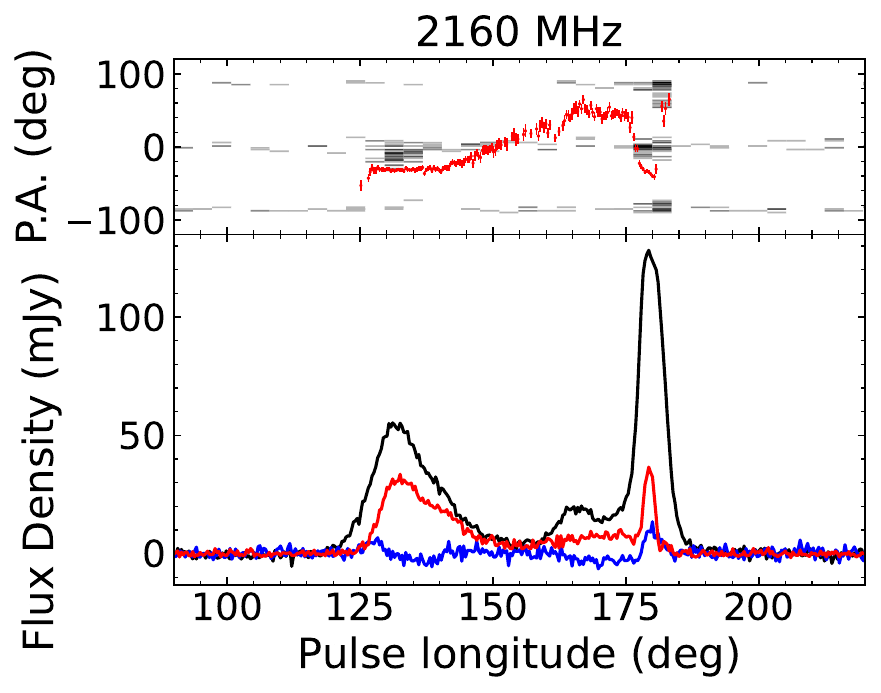}
   \includegraphics[width=0.2\columnwidth]{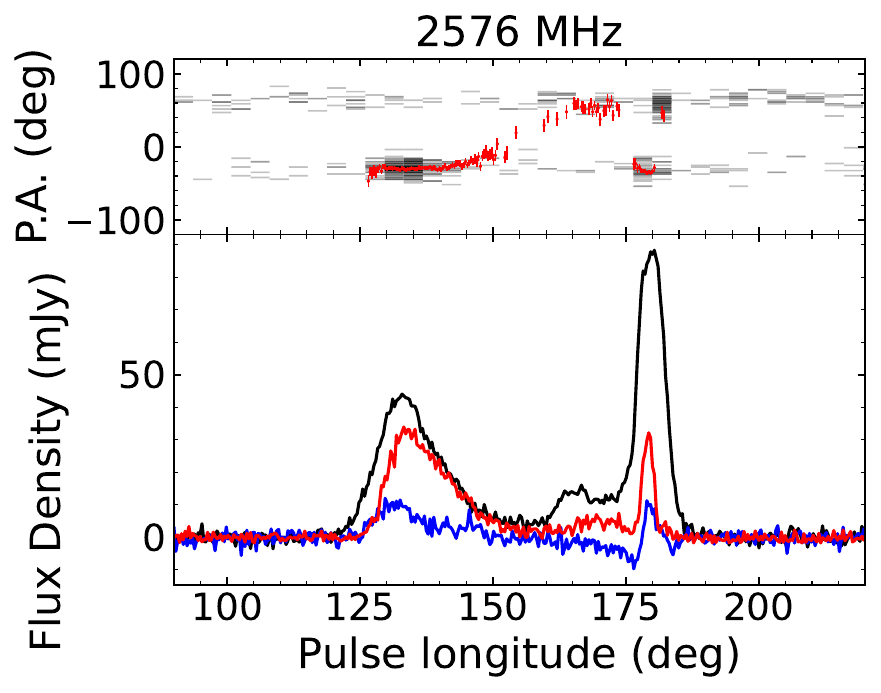}
  \includegraphics[width=0.2\columnwidth]{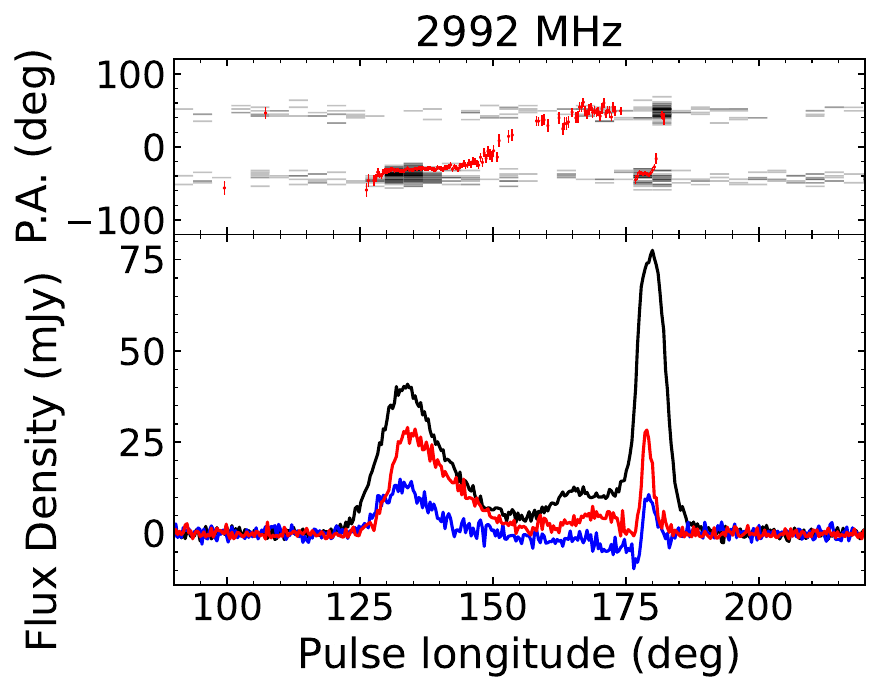}
  \includegraphics[width=0.2\columnwidth]{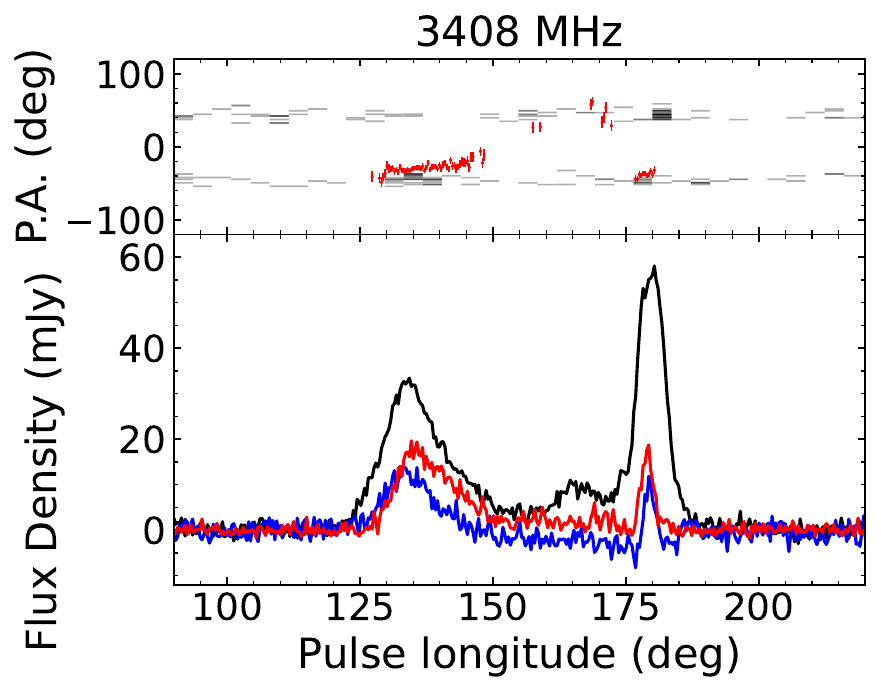}
  \includegraphics[width=0.2\columnwidth]{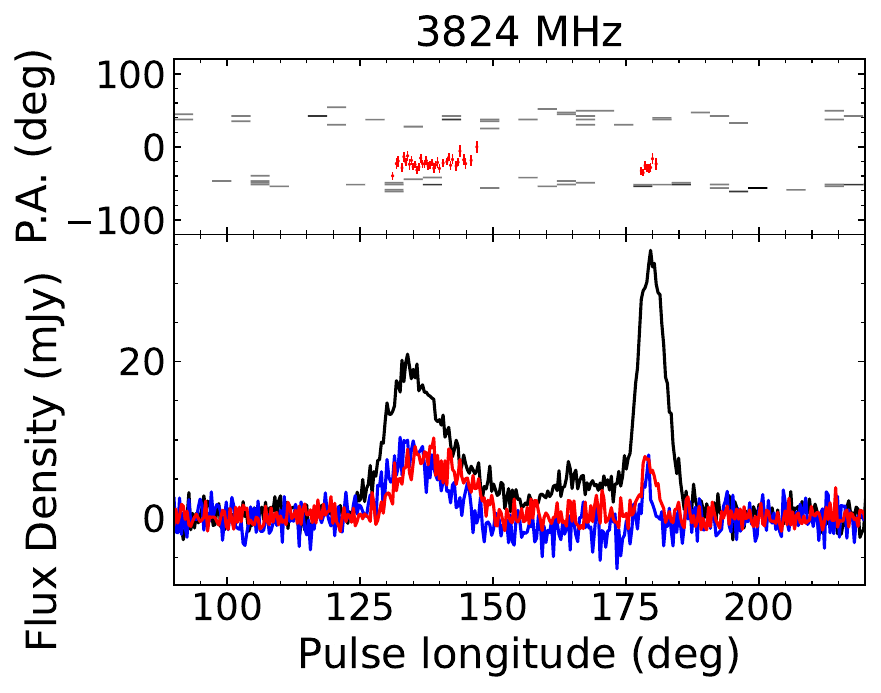}
  \centerline{Pulsar state}
   \includegraphics[width=0.2\columnwidth]{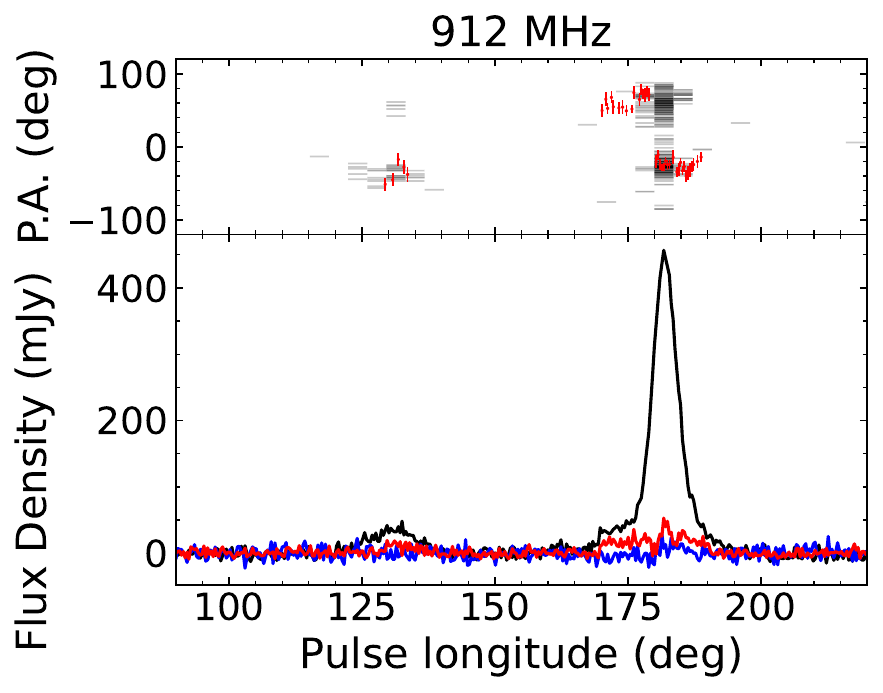}
   \includegraphics[width=0.2\columnwidth]{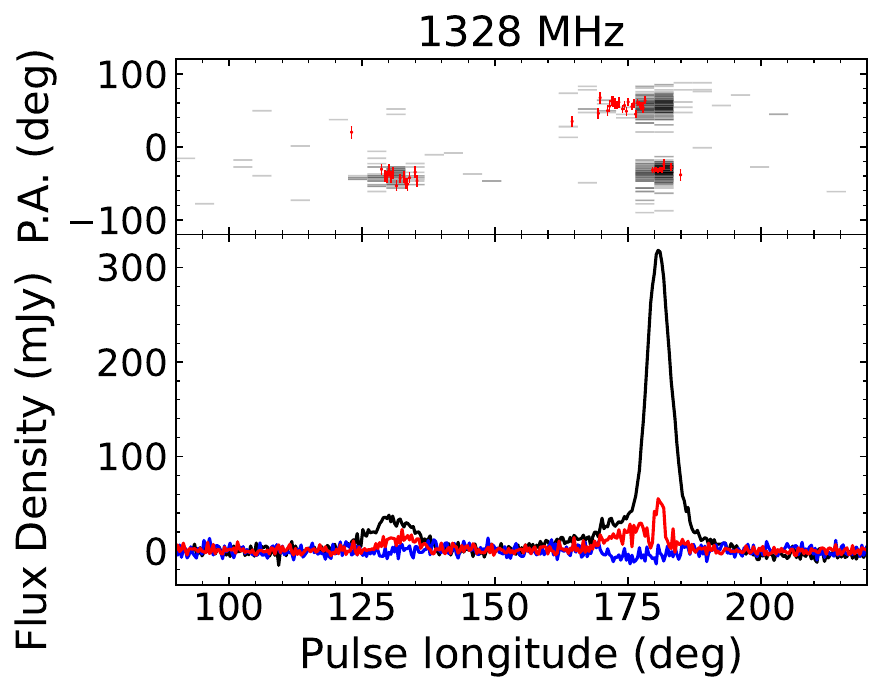}
   \includegraphics[width=0.2\columnwidth]{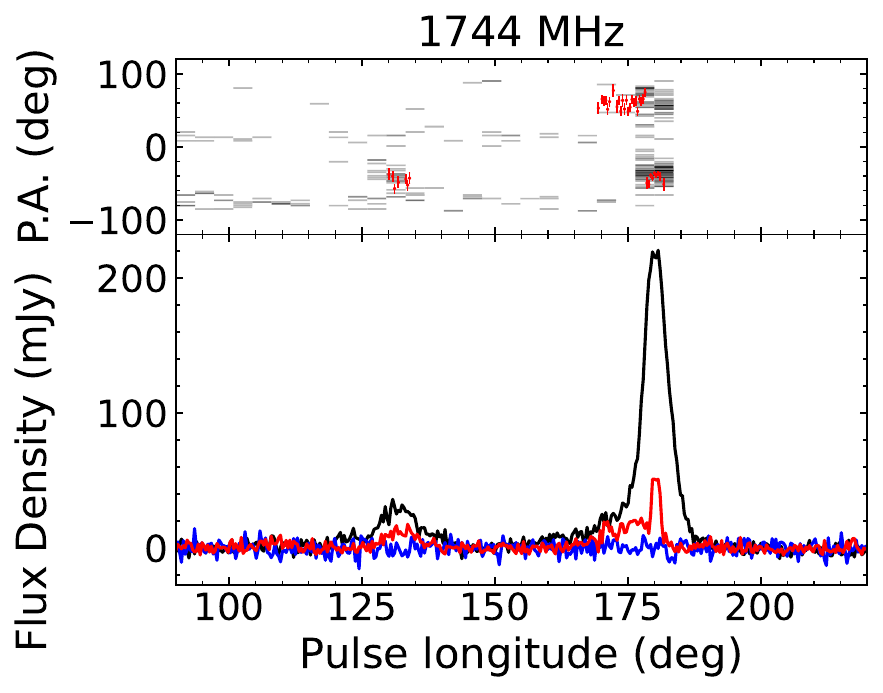}
   \includegraphics[width=0.2\columnwidth]{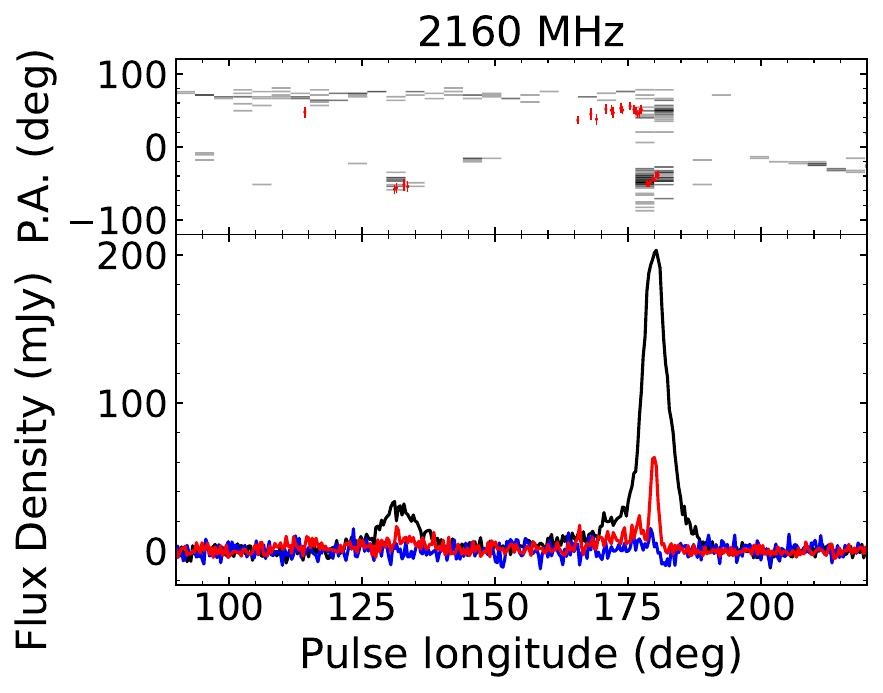}
   \includegraphics[width=0.2\columnwidth]{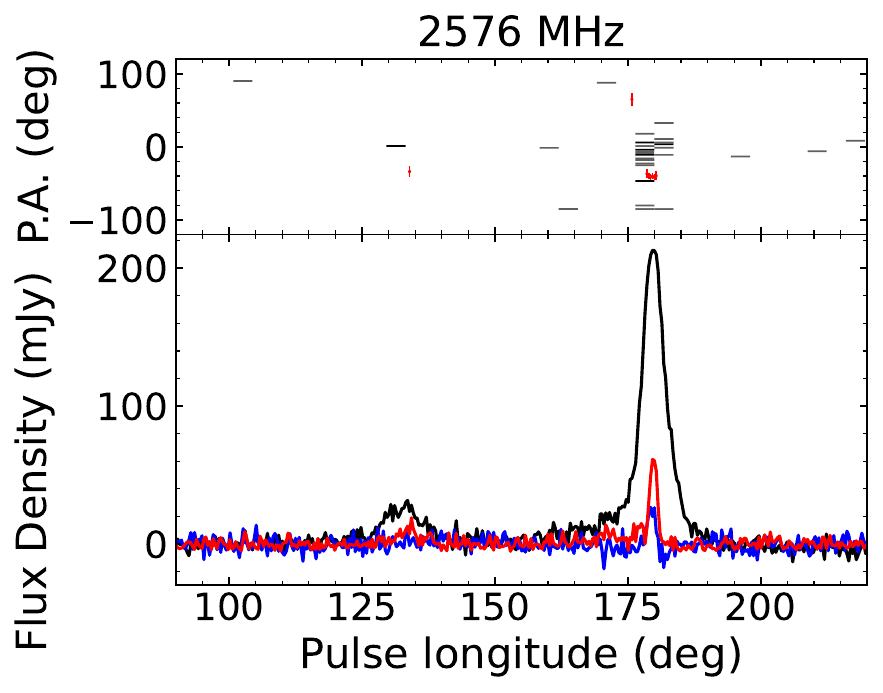}
  \includegraphics[width=0.2\columnwidth]{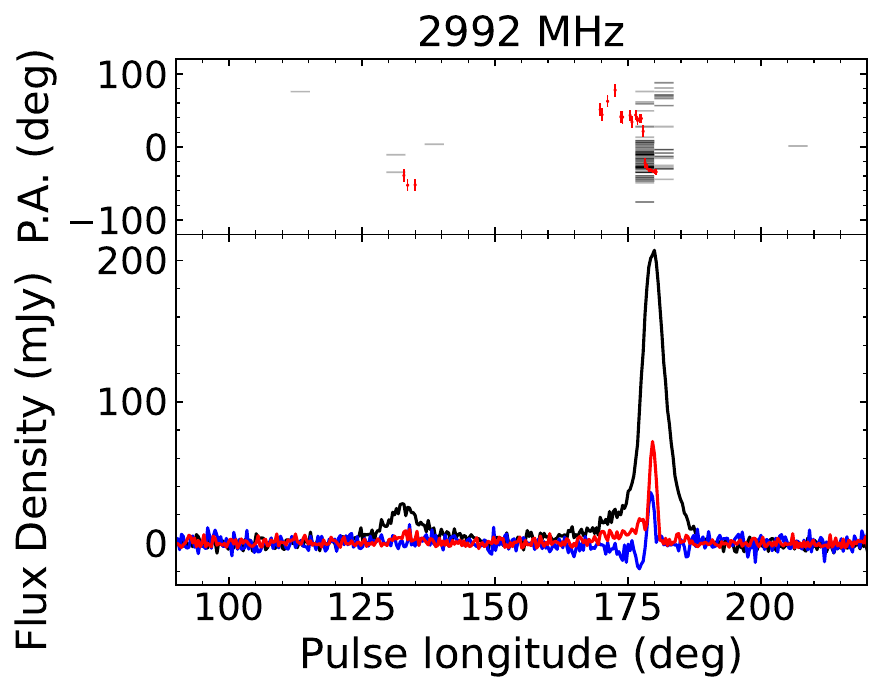}
  \includegraphics[width=0.2\columnwidth]{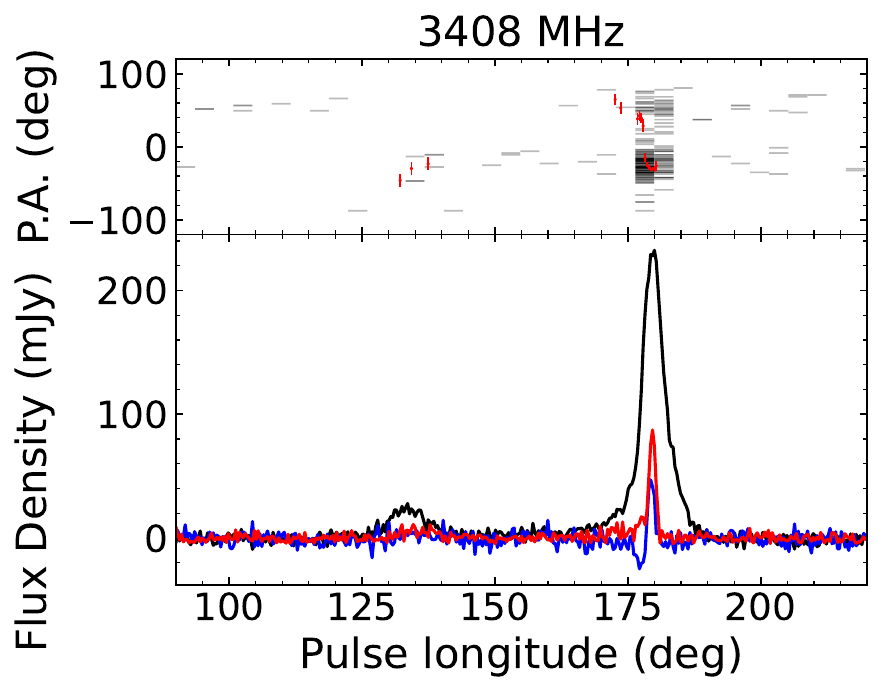}
  \includegraphics[width=0.2\columnwidth]{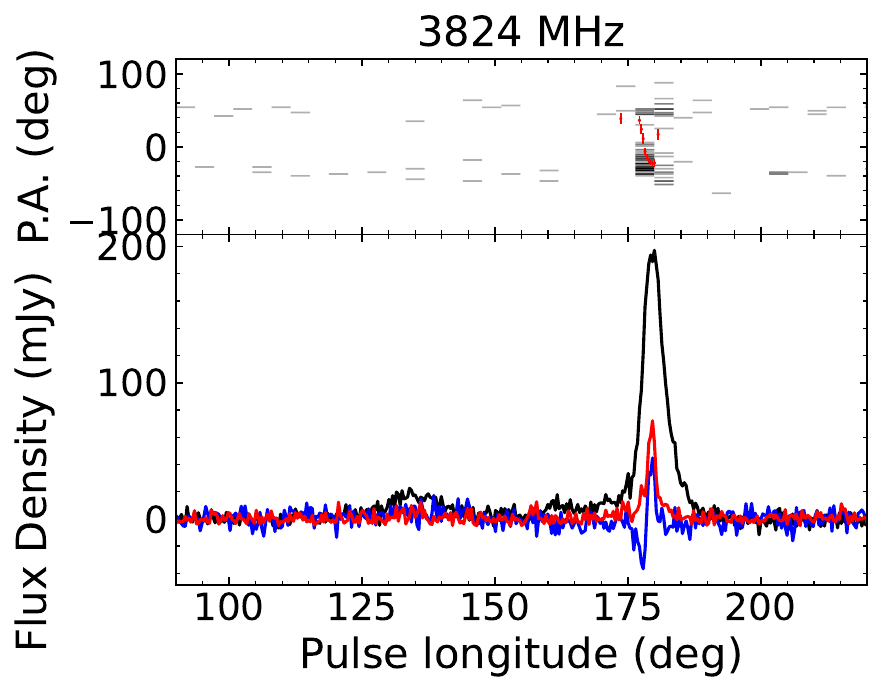}
  \centerline{RRAT state}
  \hfill
  \end{minipage}
  }
  
\subfigure[PSR J1107$-$5907 ]{
 \begin{minipage}[t]{0.92\textwidth}
  \centering
  \includegraphics[width=0.2\columnwidth]{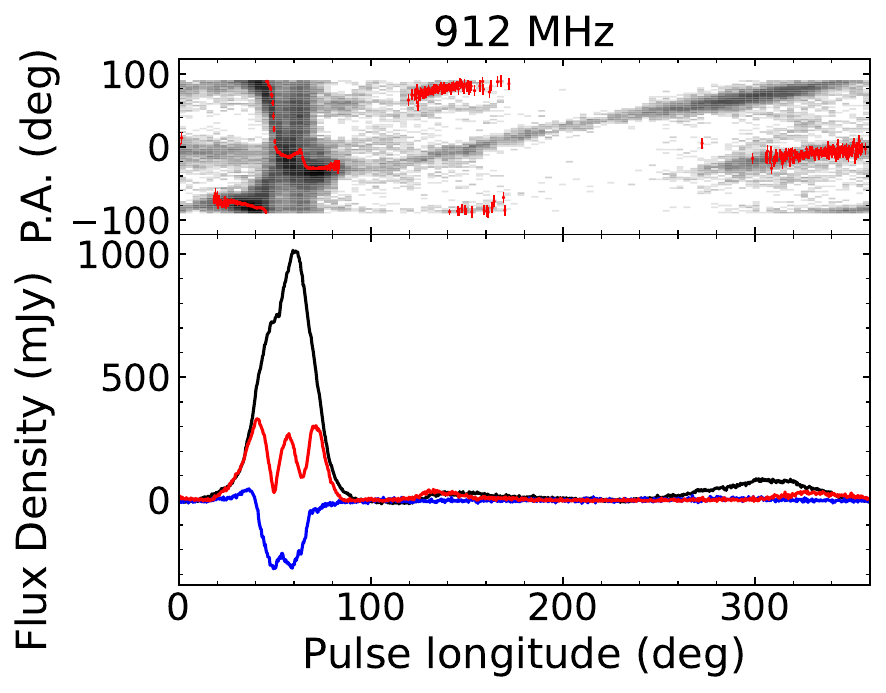}
   \includegraphics[width=0.2\columnwidth]{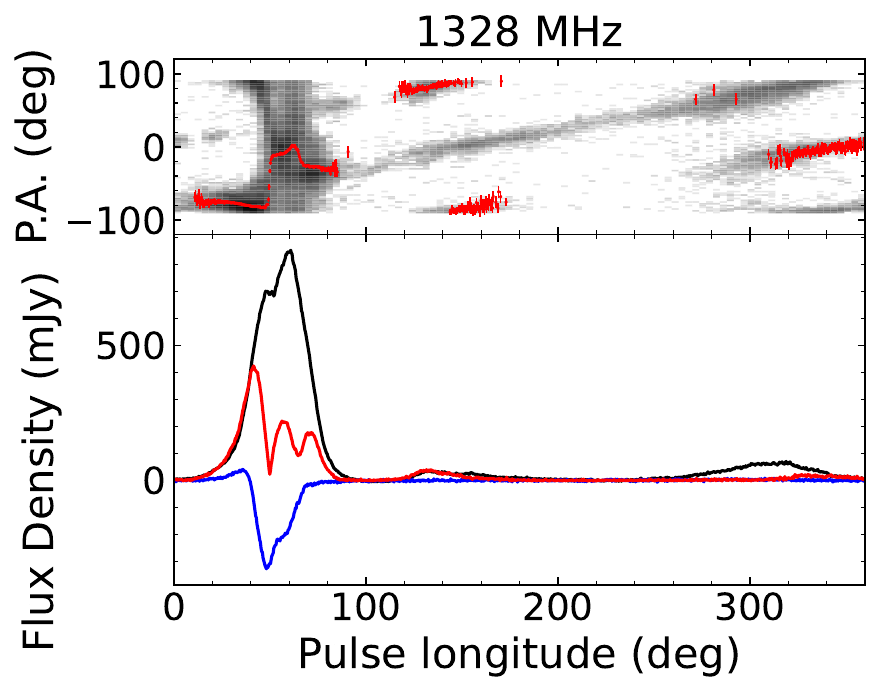}
   \includegraphics[width=0.2\columnwidth]{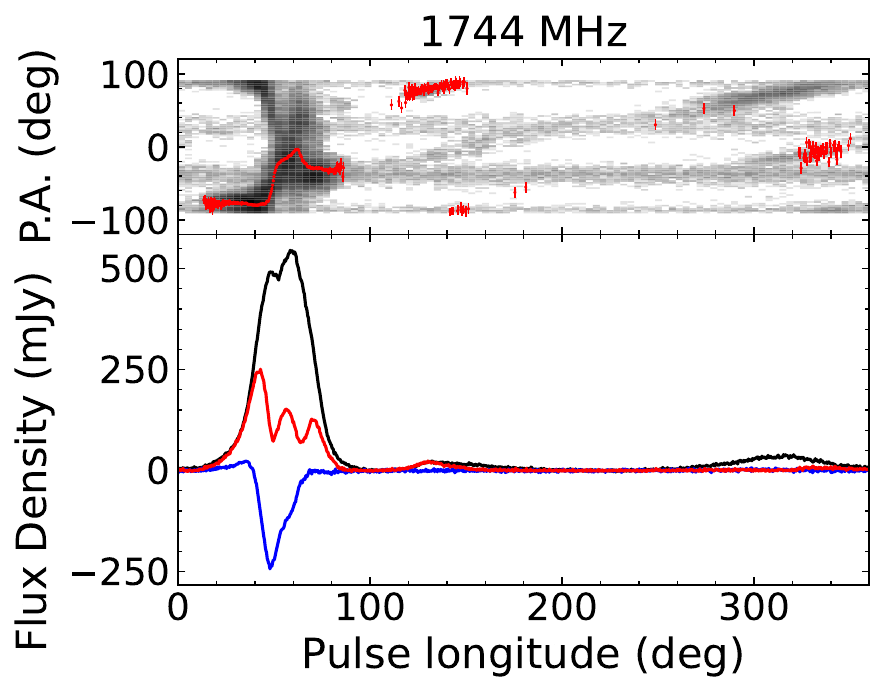}
   \includegraphics[width=0.2\columnwidth]{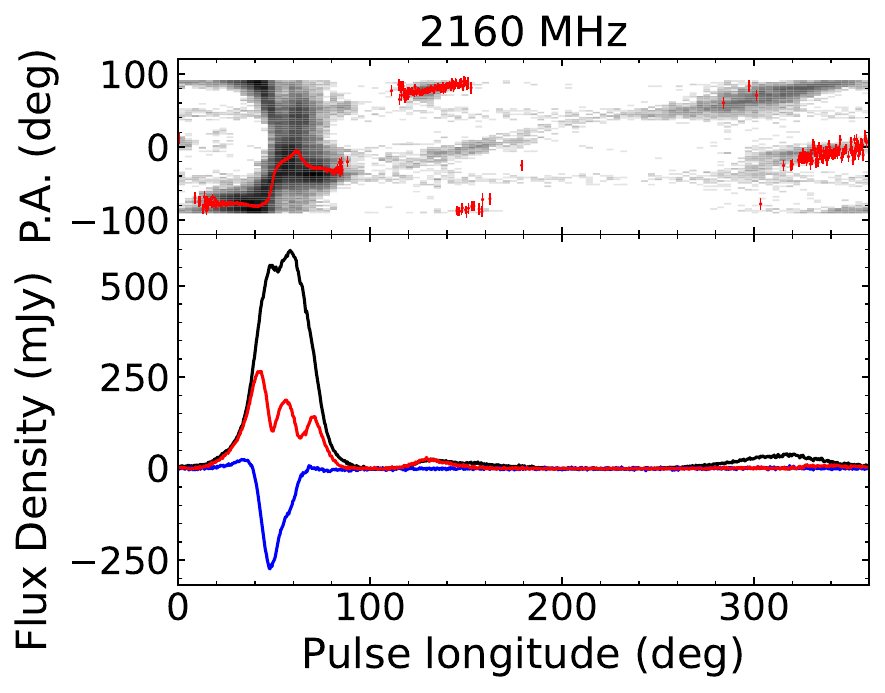}
   \includegraphics[width=0.2\columnwidth]{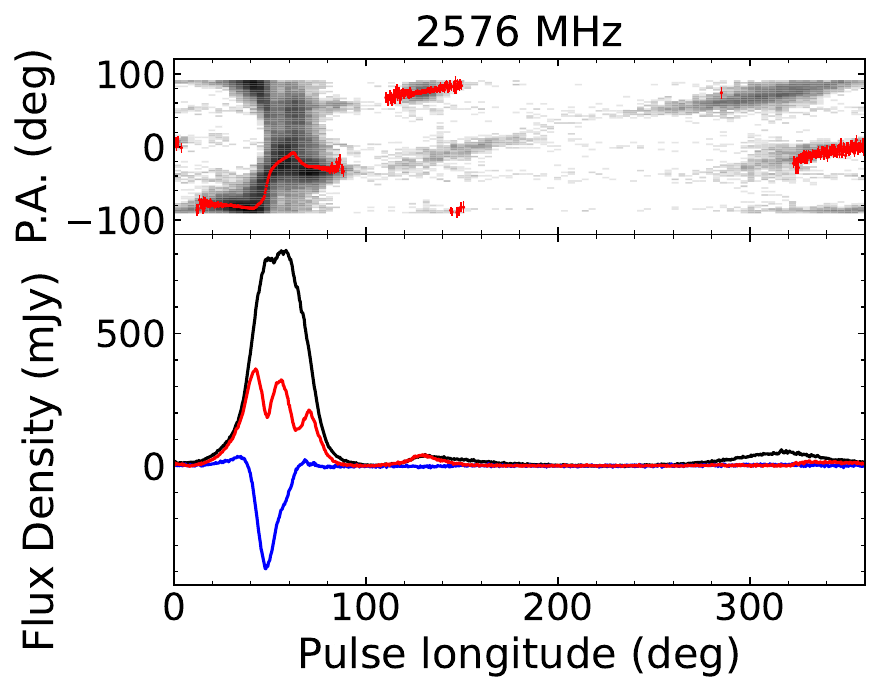}
  \includegraphics[width=0.2\columnwidth]{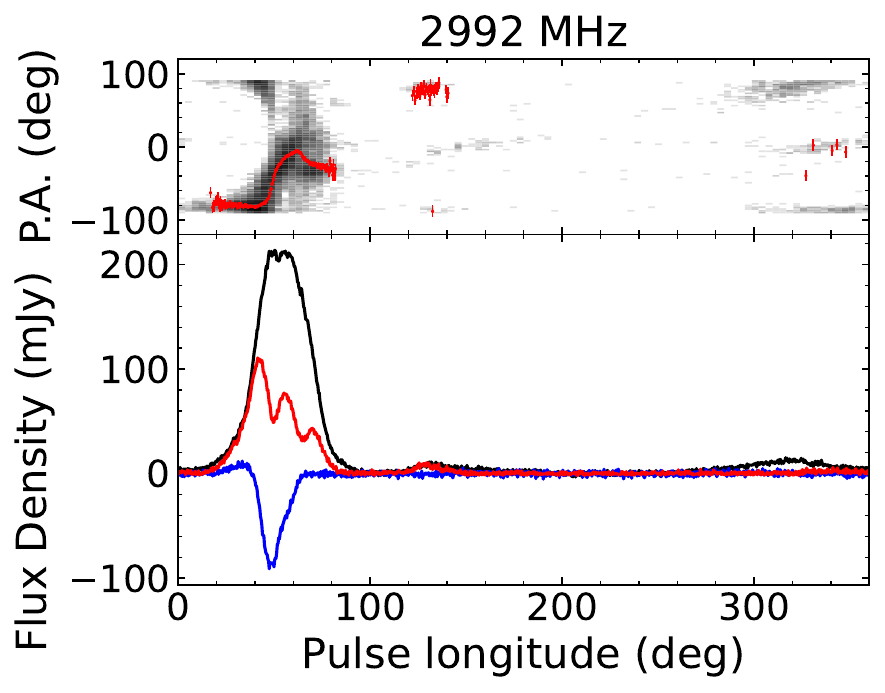}
  \includegraphics[width=0.2\columnwidth]{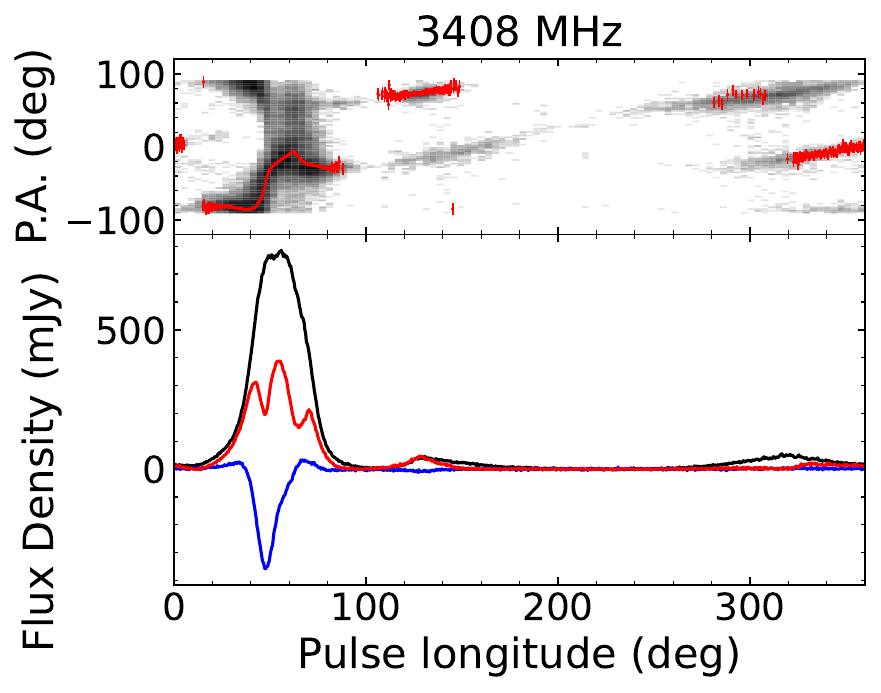}
  \includegraphics[width=0.2\columnwidth]{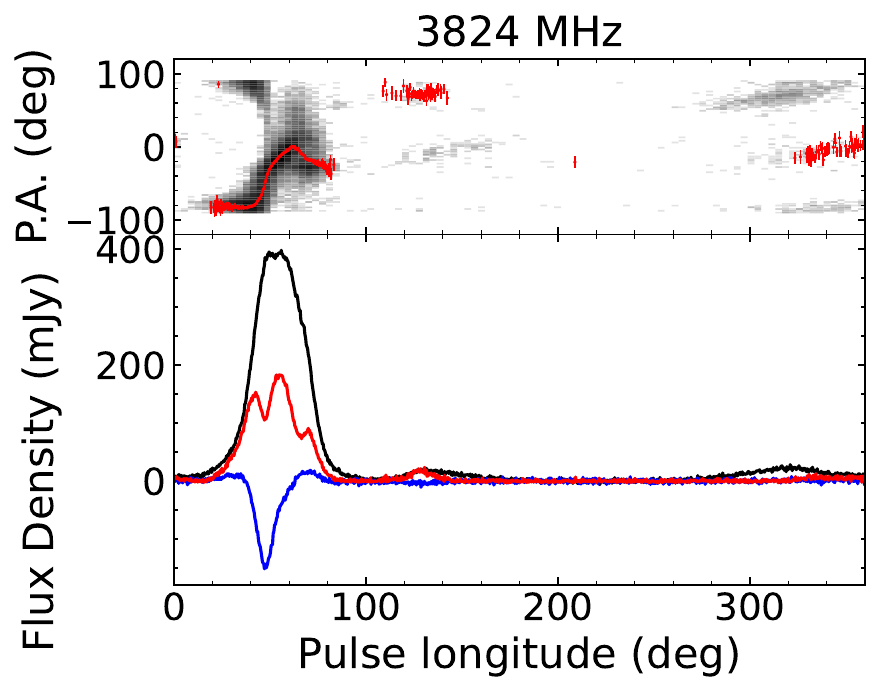}
 \centerline{Pulsar state}
  \includegraphics[width=0.2\columnwidth]{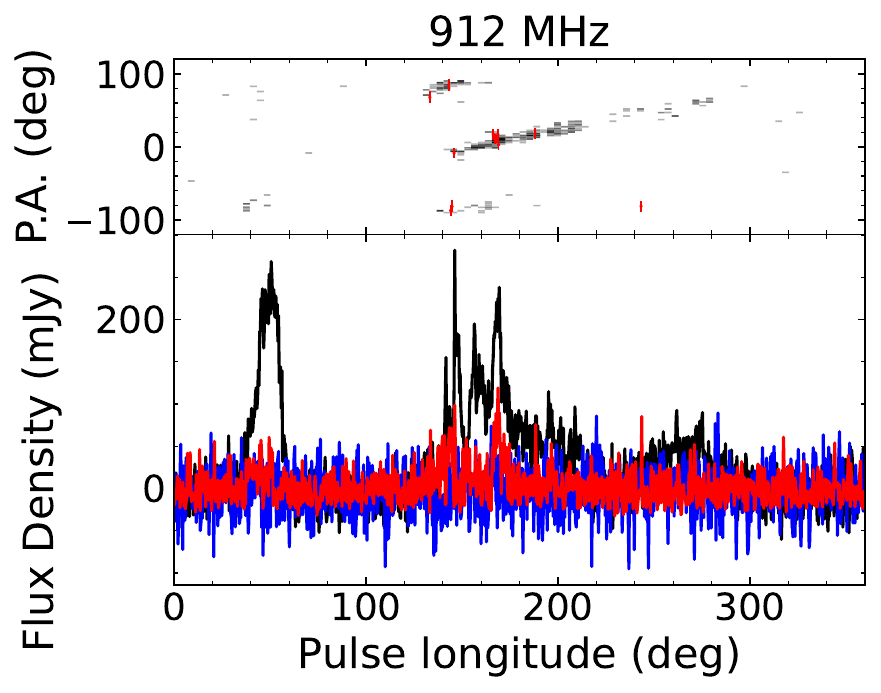}
   \includegraphics[width=0.2\columnwidth]{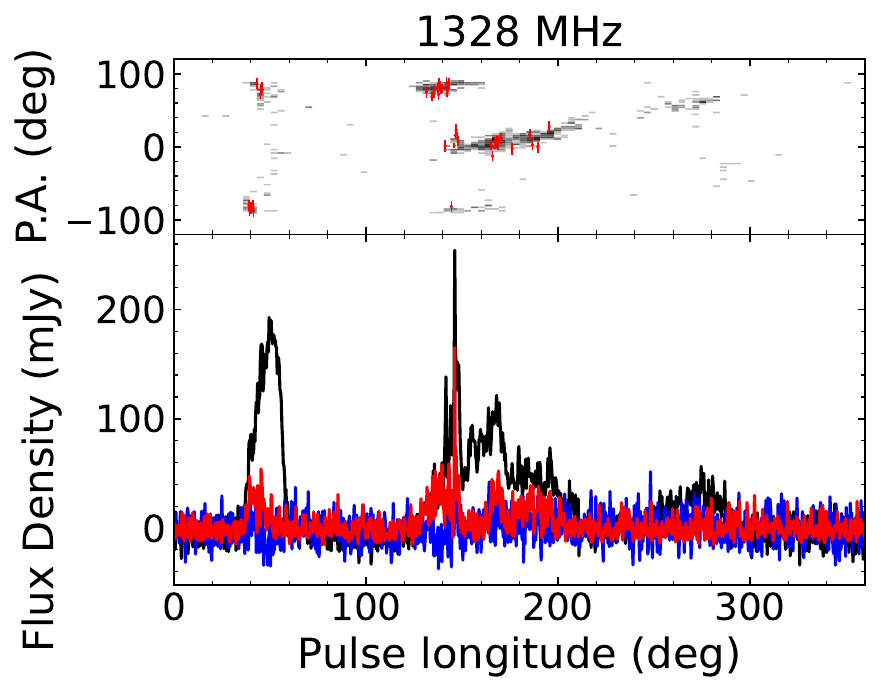}
   \includegraphics[width=0.2\columnwidth]{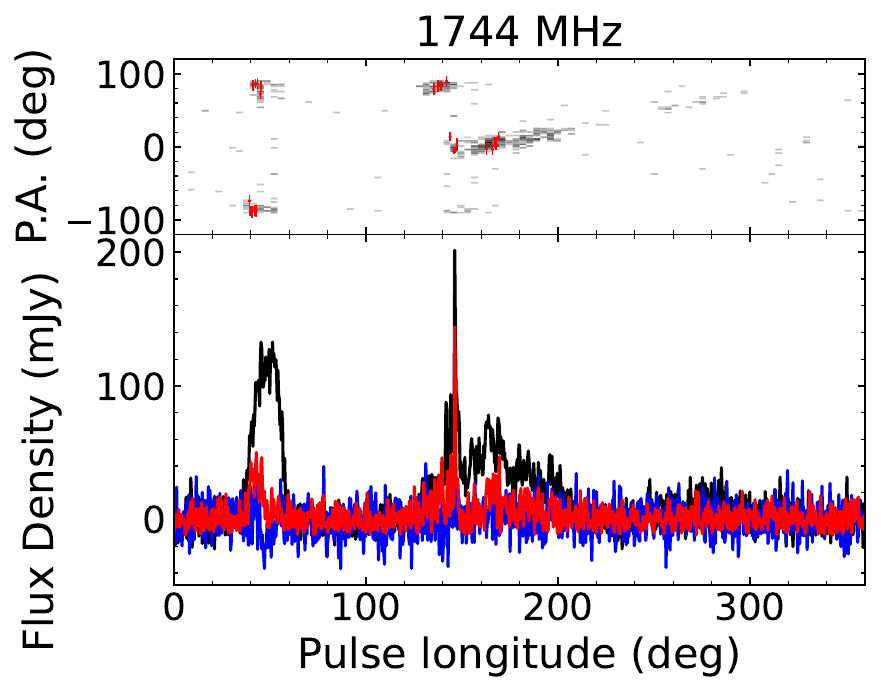}
   \includegraphics[width=0.2\columnwidth]{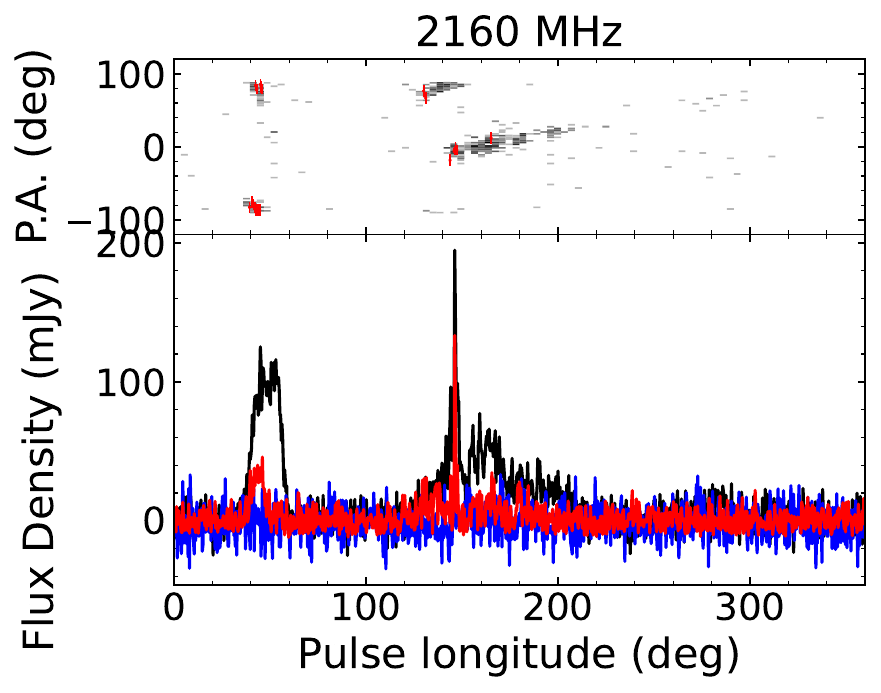}
   \includegraphics[width=0.2\columnwidth]{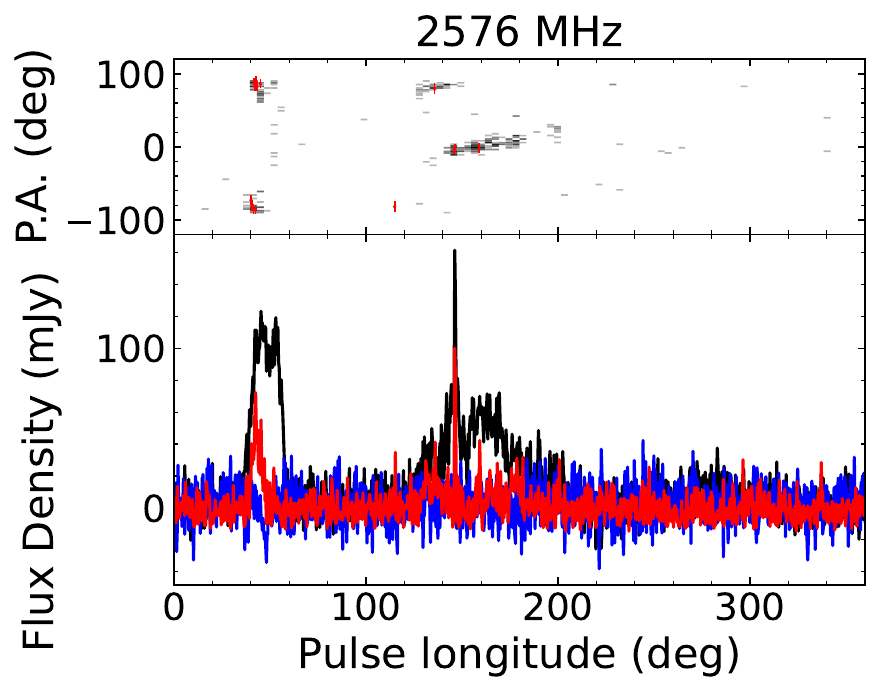}
  \includegraphics[width=0.2\columnwidth]{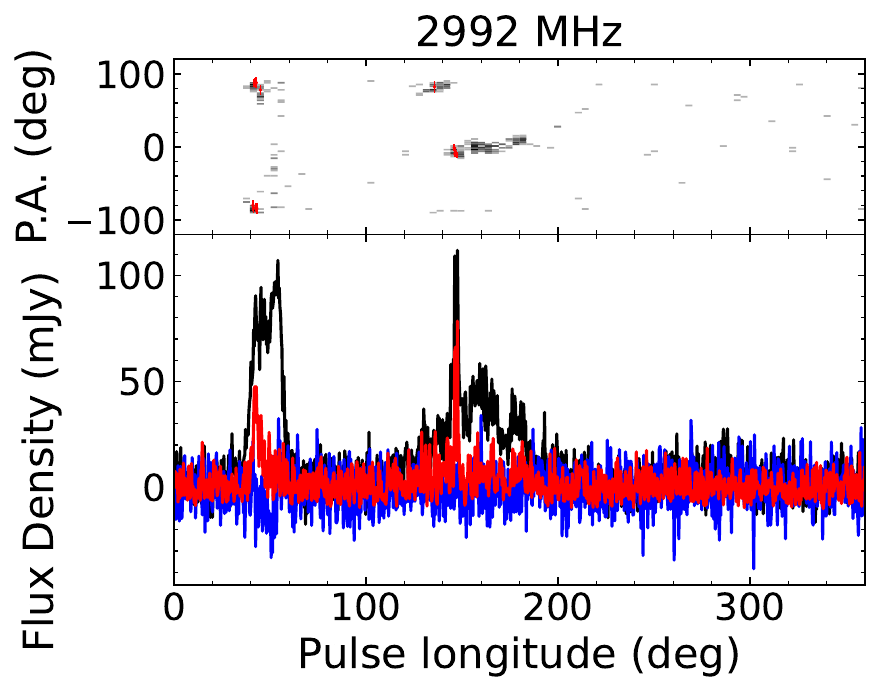}
  \includegraphics[width=0.2\columnwidth]{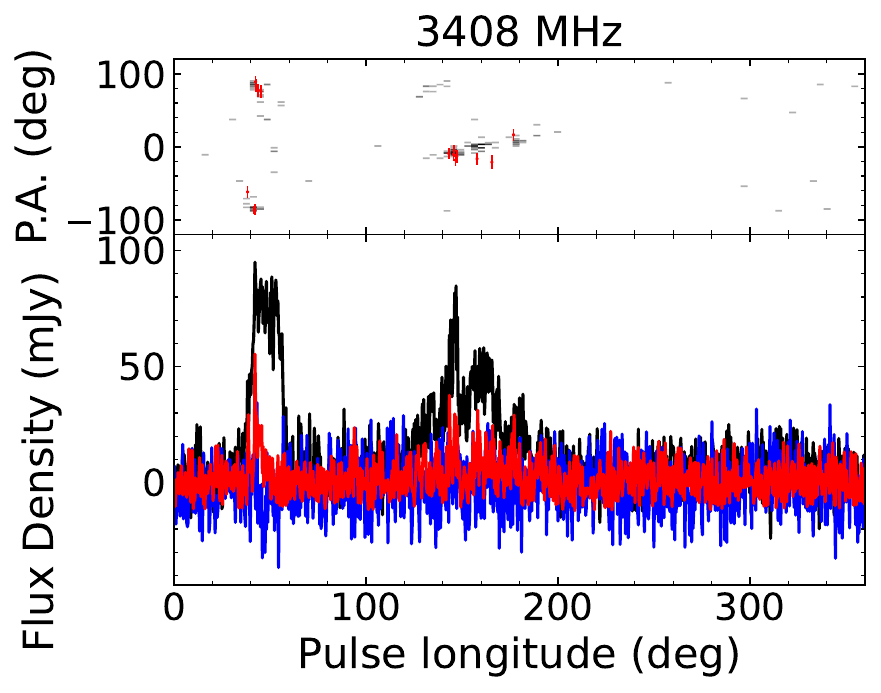}
  \includegraphics[width=0.2\columnwidth]{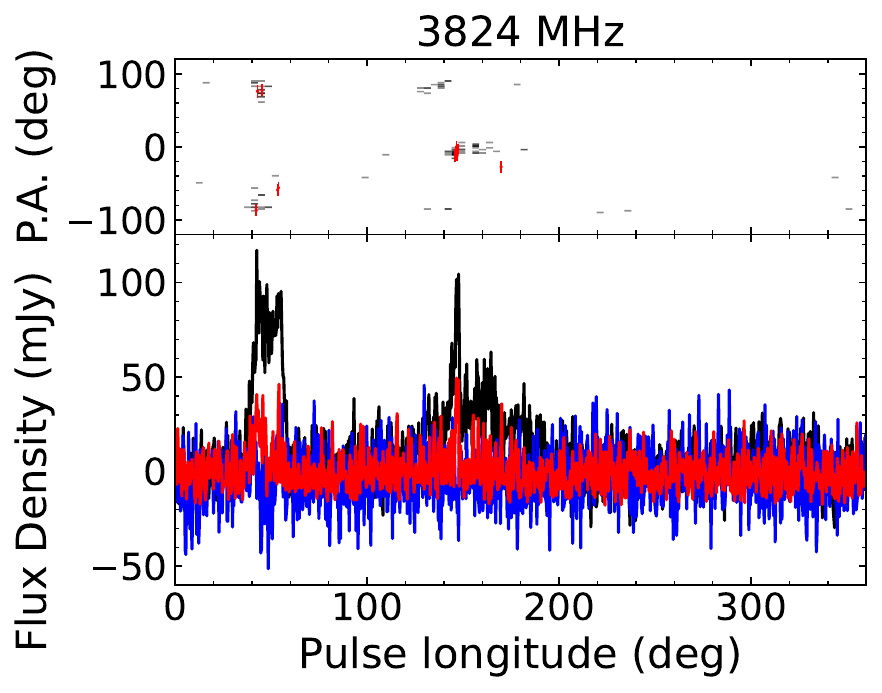}
  \centerline{RRAT state}
  \hfill
 \end{minipage}
  } 
 \caption{The polarization profiles of 8 sub-bands with a bandwidth of 416\,MHz for the pulsar and RRAT states for PSR J0941$-$39 (panel (a)) and PSR J1107$-$5907 (panel (b)). The labels are the same as in Figure~\ref{fig:1pa}.}
 \label{fig:1poln}
 \end{figure*}

 \begin{figure*}
\centering
\subfigure[PSR J0941$-$39]{
 \includegraphics[width=1\columnwidth]{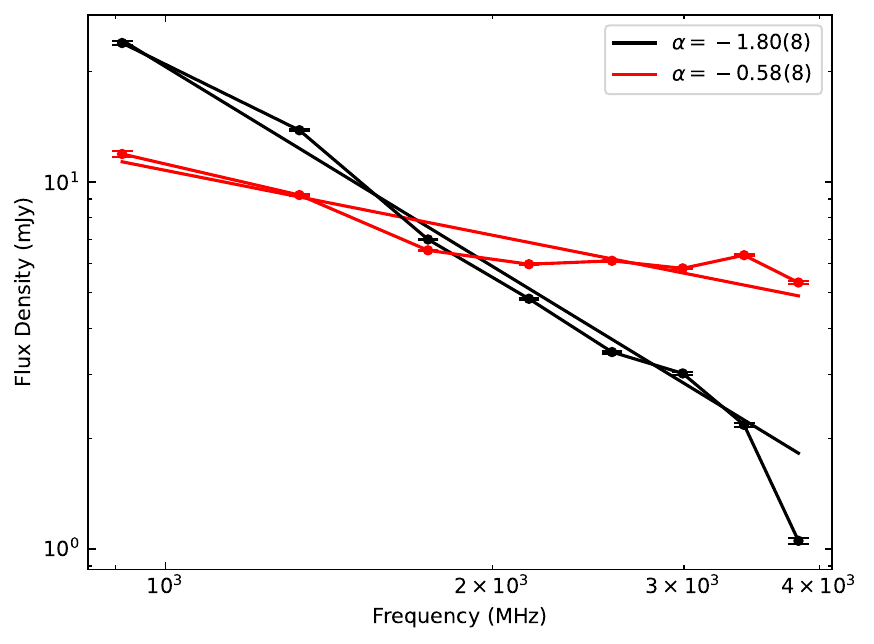}
 }
   \subfigure[PSR J1107$-$5907]{
  \includegraphics[width=1\columnwidth]{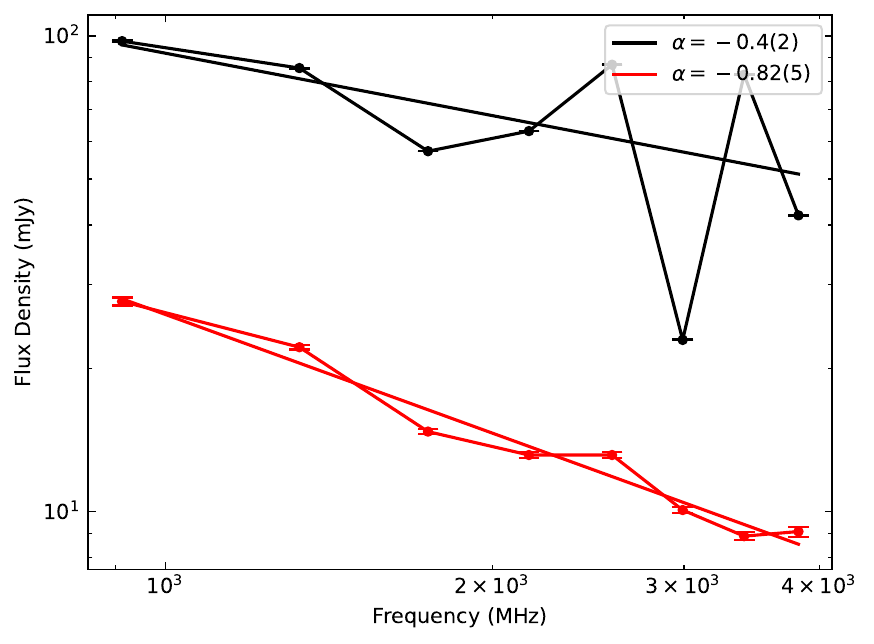}
  }
 \caption{Flux density as a function of the frequency of pulsar (black points with solid line) and RRAT (red points with solid line) states for PSR J0941$-$39 (panel (a)) and PSR J1107$-$5907 (panel (b)). The solid lines are for the power-law fitting results of different states of each pulsar.}
 \label{fig:1flux}
 \end{figure*}

\begin{table*}
\renewcommand\arraystretch{1.3}
\centering
\caption{Flux density and polarization parameters of the two states of PSR J0941$-$39 and PSR J1107$-$5907 at 8 sub-bands.}
\label{mutipoln}
\begin{tabular}{ccccccccccc}
\hline
\multirow{2}{*}{PSR}  & \multirow{2}{*}{State} &  & \multicolumn{8}{c}{Central Observation Frequency (MHz)} \\
&  &  & 912 & 1328 & 1744 & 2160 & 2576 & 2992 & 3408 & 3824\\ 
\hline
\multirow{8}{*}{J0941$-$39}  & \multirow{4}{*}{Pulsar} & $f_{\rm L }\,(\%)$ & $37.7(1)$ & $38.1(3)$ & $44.6(6)$ & $37.0(1)$ & $42.4(4)$  & $39.2(6)$ & $31.5(2) $ & $27(2)$ \\
&  & $f_{\rm C}\,(\%)$ & $-5.4(3)$ & $-3.3(3)$ & $-3(1)$ & $1(2)$ & $5.5(1)$ & $9(2)$ & $8(2)$ & $12(3)$ \\
&  &  $f_{\rm \lvert C \lvert}\,(\%)$ &$9.3(3)$ & $8.7(3)$ & $11(1)$ & $11(2)$ & $19.1(1)$ & $22(2)$ & $29(2)$ & $38(3)$ \\
&  & $S$\,(mJy) & $23.9(2)$ & $13.8(1)$ & $6.9(1)$ & $4.8(1)$ & $3.4(1)$ & $3.0(1)$ & $2.1(1)$ & $1.0(1)$\\  
& \multirow{4}{*}{RRAT} & $f_{\rm L}\,(\%) $& $14(1)$ & $19(2)$ &$20(2)$ & $18(2)$ & $13.2(5)$ & $14.4(5)$ & $14(1)$ & $14.5(1)$ \\
&  & $f_{\rm C}\,(\%)$& $-1(4)$ & $-1(1)$ & $0(1)$ & $1.8(4)$ & $0(3)$ & $0(2)$ & $-3(3)$ & $2(2)$ \\
&  & $f_{\rm \lvert C \lvert}$\,(\%) & $19(4)$ &$16(1)$ & $19(1)$ & $20.4(4)$ & $24(3)$ & $23(2)$ &$27(3)$ &$34(2)$ \\
&  & $S$\,(mJy) & $11.9(2)$ & $9.2(1)$ & $6.5(1)$ & $5.9(1)$ & $6.1(1)$ & $5.8(1)$ & $6.3(1)$ & $5.3(1)$ \\
\multirow{8}{*}{J1107$-$5907} & \multirow{4}{*}{Pulsar} & $f_{\rm L}\,(\%)$ & $35.8(4)$ &$38.3(1)$ & $36(1)$ & $38.2(1)$ & $41(1)$ & $39(2)$ & $44(3)$ & $39.7(5)$ \\
&  & $f_{\rm C}\,(\%)$ & $-15(2)$ & $-16(2)$  & $-15(1)$ & $-14.2(1)$ & $-13(2)$ &$ -12.5(1)$ & $-13(2)$ & $-10(3)$ \\
&  &  $f_{\rm \lvert C \lvert}$\,(\%) & $21(2)$ & $20(2)$ & $20(1)$ & $19.7(1)$& $19(2)$ & $20.3(1)$ & $18(2)$ & $18(3)$\\
&  & $S$\,(mJy) & $97.4(4)$ & $85.5(2)$ & $57.2(1)$ & $63.0(1)$ & $86.9(1)$ & $22.9(1)$ & $82.7(1)$ & $41.9(1)$ \\
& \multirow{4}{*}{RRAT} & $f_{\rm L }$\,(\%)& $7(2)$ & $13(1)$ & $18(2)$ & $17(2)$ & $15(3)$ & $17.2(1) $ & $11(3)$ & $15.9(5)$  \\
& & $f_{\rm C}\,(\%)$& $-3(4)$ & $-3.6(5)$ & $-3(2)$ & $0(4)$ & $0.1(6)$ & $-5(2)$ & $-7(6)$ & $-7.3(2)$ \\
&  & $f_{\rm \lvert C \lvert}$\,(\%) & $20(4)$ & $13.5(5)$ & $16(2)$ &$15(4)$ & $14.2(6)$ & $18(2)$ & $18(6)$ & $24.9(2)$ \\
&  & $S$\,(mJy) & $27.6(5)$ & $22.1(2)$ & $14.7(1)$ & $13.1(1)$ & $13.1(1)$ & $10.0(1)$ & $8.8(1)$ & $9.0(2)$ \\
\hline
 \end{tabular}
\end{table*}

 \begin{figure*}
\centering
 \includegraphics[width=\textwidth]{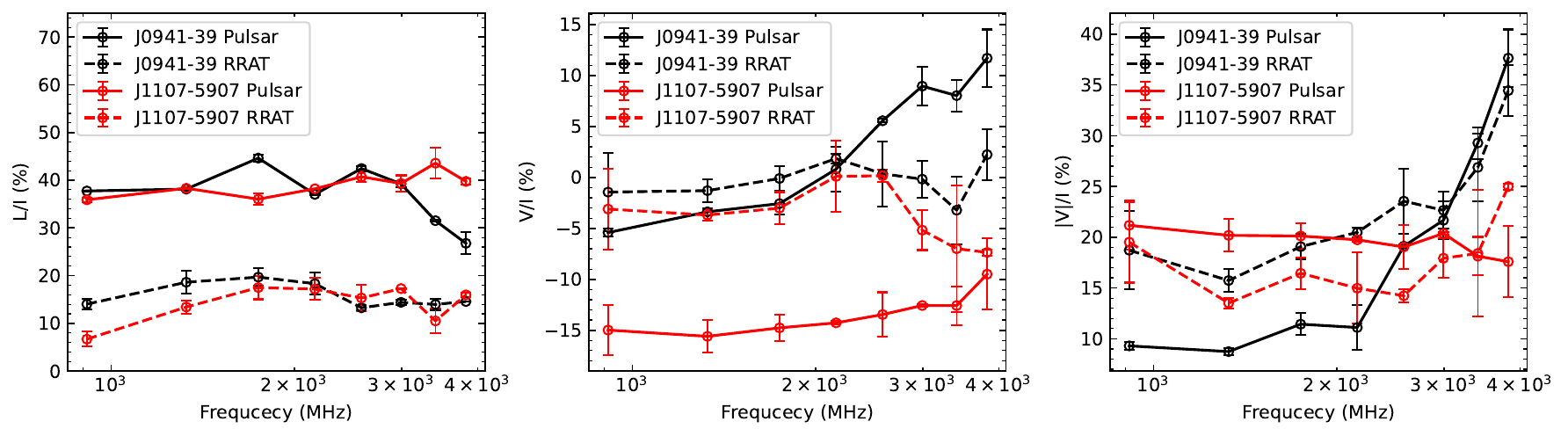}
 \caption{Summary of fractional linear (left panel), circular (middle panel), and absolute circular polarization (right panel) versus frequency for PSR J0941$-$39 (black lines) and PSR J1107$-$5907 (red lines). The solid lines, dashed lines are for the pulsar and RRAT states respectively.}
 \label{fig:allpoln}
 \end{figure*}

\subsubsection{PSR J1107$-$5907}

We present a long-duration observation of PSR J1107$-$5907 lasting approximately 9.5 hours, during which both the pulsar state and RRAT state were detected. The RRAT state was observed three times, with durations of 27\,minutes, 170\,minutes and 333\,minutes respectively. The pulsar state was detected twice, with durations of 9\,minutes and 27\,minutes respectively. Panel (b) of Figure~\ref{fig:1state} shows the single-pulse stacks for both the pulsar and RRAT states, revealing significant pulse-to-pulse variations in the pulsar state and sporadic single pulses in the RRAT state. Following the method used for PSR J0941$-$39, we identified a total of 281 single pulses with $S/N_{p} >7$ during the RRAT state of PSR J1107$-$5907.

The frequency-averaged pulse profiles of the pulsar and RRAT states are presented in panel (b) of Figure~\ref{fig:1pa}. The pulse emission from PSR J1107$-$5907 spans the entire pulse profile, consistent with previous results (e.g.~\cite{yws+2014}). In the pulsar state, the average profile exhibits a prominent component along with two weaker components. In contrast, during the RRAT state, the average profile also exhibits three components but with stronger first and second components. For the pulsar state, we find that $f_{\rm L}$ is measured to be $40(2)\%$, $f_{\rm C}$ is measured to be $-14(1)\%$, and $f_{\rm \lvert C \lvert}$ is measured to be $17(1)\%$. For the RRAT state, these values are determined as $19(7)\%$, $-7(4)\%$, and $28(4)\%$, respectively. The observed PA swings in both pulsar and RRAT states exhibit complex variations that deviate from the expected S-shaped pattern. Three OPMs have been identified at approximately 100\,deg, 180\,deg and 270\,deg pulse phases within the average profile of pulsar state. For the average profile of the RRAT state, it is hard to determine where exactly OPM occurs due to limited S/N.

The pulse fluence distributions for the pulsar state and RRAT state are presented in panel (b) of Figure~\ref{fig:1ehist}, while the corresponding fitting parameters are shown in Table~\ref{fit}. For the pulsar state, a single function is insufficient to adequately fit the fluence distribution, which aligns with the results reported by~\cite{mtb+2018}. 
We used different functions to fit different components of the fluence distribution. 
Our analysis suggests that the first component of the fluence distribution in the pulsar state most likely follows a log-normal distribution. For the high fluence tail, as mentioned by~\cite{mtb+2018}, we only performed fittings on pulses with a normalized fluence $F\geq 4 \langle F \rangle$. Subsequently, both power-law and log-normal functions were used to fit the fluence distribution. Our results indicate that log-normal fitting yielded better results than power-law fitting, which agrees with previous observations at 154\,MHz and 835\,MHz of~\citet{mtb+2018}. For the RRAT state of PSR J1107$-$5907, the Gaussian fitting yielded better results than log-normal fitting of pulse fluence, however, this result is constrained by the limited number of single pulses.

The polarization profiles of the 20 brightest single pulses in the RRAT state of PSR J1107$-$5907 are presented in Figure~\ref{fig:1singlepulses}. The W50 pulse widths range from 0.34\,ms to 4.39\,ms. Significant variations are observed among the profiles of different single pulses, with some exhibiting multiple components. The $f_{\rm L}$ values for these single pulses range from 28\% to 100\%, while the $f_{\rm C}$ is much smaller, ranging from $-28$\% to 28\%. Some single pulses exhibit flat PA swings that can vary significantly between them. The peak intensity of the brightest single pulse in the RRAT state is approximately 24 times stronger than that of the average pulse profile in the pulsar state, with the peak intensity of 15 Jy and the fluence of 25 Jy ms.

 \subsection{Polarization profiles at different frequencies}

To investigate the frequency-dependent evolution of polarization profiles for the two pulsars in different states, we divided the entire bandwidth into 8 sub-bands of 416\,MHz width. The polarization profiles of these two pulsars within these 8 sub-bands are illustrated in Figure~\ref{fig:1poln}, while Table~\ref{mutipoln} presents the values of $f_{\rm L}$, $f_{\rm C}$ and $f_{\rm \lvert C \lvert}$.

The polarization profiles of both pulsars exhibit significant variations with increasing frequency. {For PSR J0941$-$39, in the pulsar state, as the frequency increases}, there is a noticeable increase in strength for the first component compared to the third component. However, during the RRAT state, the first component becomes much weaker than the third component and almost disappears at high frequencies. By combining this observation with rotating vector model (RVM) fitting (see Section 3.3 for more detail), we can conclude that both the first and third components are associated with cone emission. Furthermore, we have measured the flux density of different states at various frequencies (Figure~\ref{fig:1flux}), and these values are provided in Table~\ref{mutipoln}. We then obtained spectral indices through power-law fitting (solid lines in Figure~\ref{fig:1flux}). The spectral index for the pulsar state is $-1.80(8)$, which is steeper than that observed during RRAT state of $-0.58(8)$ (panel (a) of Figure~\ref{fig:1flux}). For PSR J1107$-$5907, during its pulsar state, it is evident that the first component is significantly stronger than both second and third components, however, during its RRAT state, it's observed that second components become stronger instead. The spectral index for the pulsar state is $-0.4(2)$, which is flatter compared to an index of $-0.82(5)$ observed in its RRAT state.

The frequency dependence of $f_{\rm L}$, $f_{\rm C}$ and $f_{\rm \lvert C \lvert}$ for different states in PSR J0941$-$39 (panel (a)) and PSR J1107$-$5907 (panel (b)) is depicted in Figure~\ref{fig:allpoln}. These two pulsars exhibit varying degrees of linear and circular polarization across different states. Both $f_{\rm L}$, as well as $f_{\rm C}$ and $f_{\rm \lvert C \lvert}$, exhibit nonmonotonic variations with increasing frequency. However, a general tendency is observed where $f_{\rm L}$ tends to decrease while $f_{\rm C}$ increases with increasing frequencies. For the PA swings, both PSR J0941$-$39 and PSR J1107$-$5907 exhibit similar shapes at different frequencies.

\begin{figure*}
\centering
 \subfigure[Pulsar state]{
\includegraphics[width=1\columnwidth]{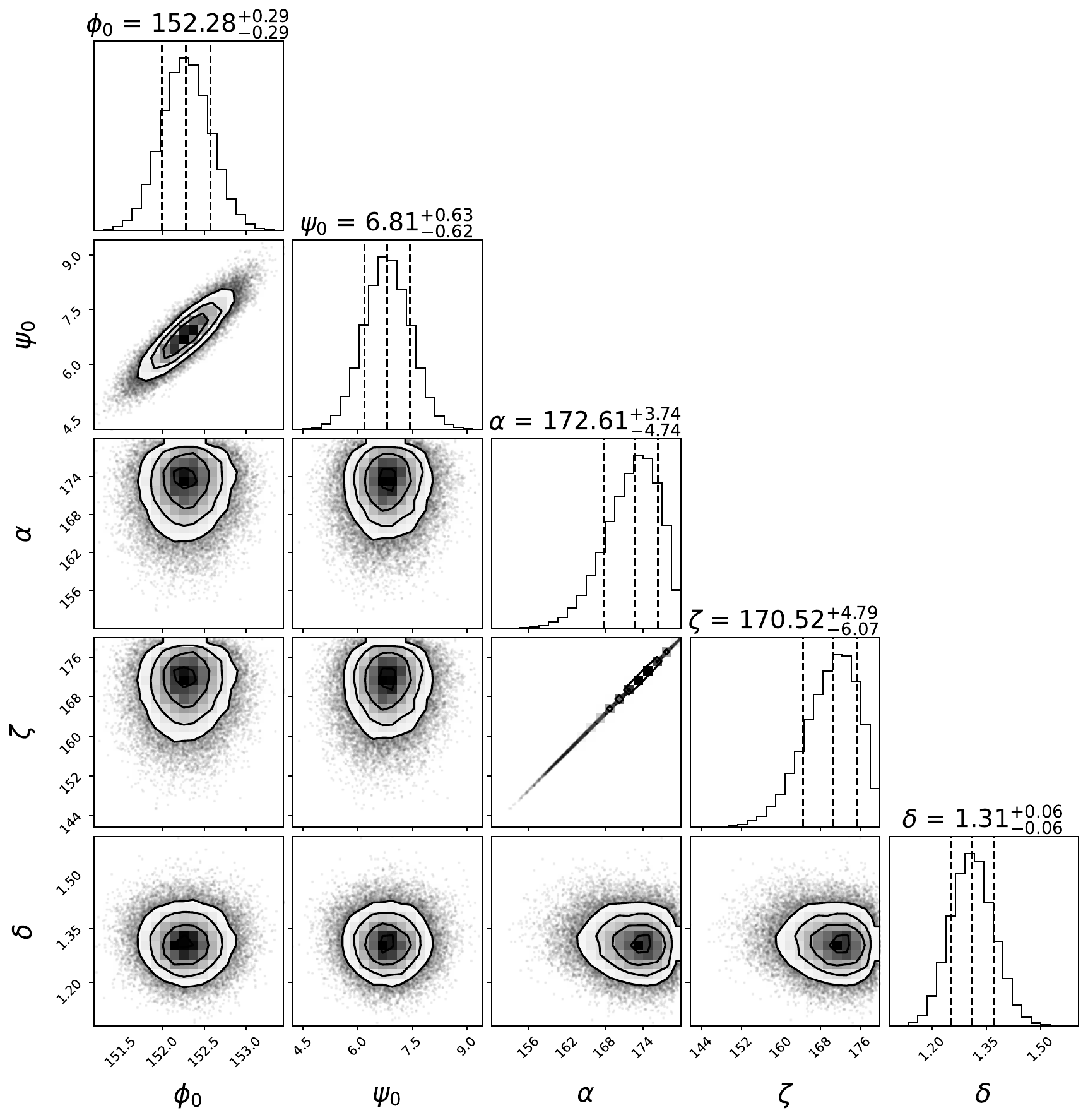}}
\subfigure[RRAT state]{
\includegraphics[width=1\columnwidth]
 {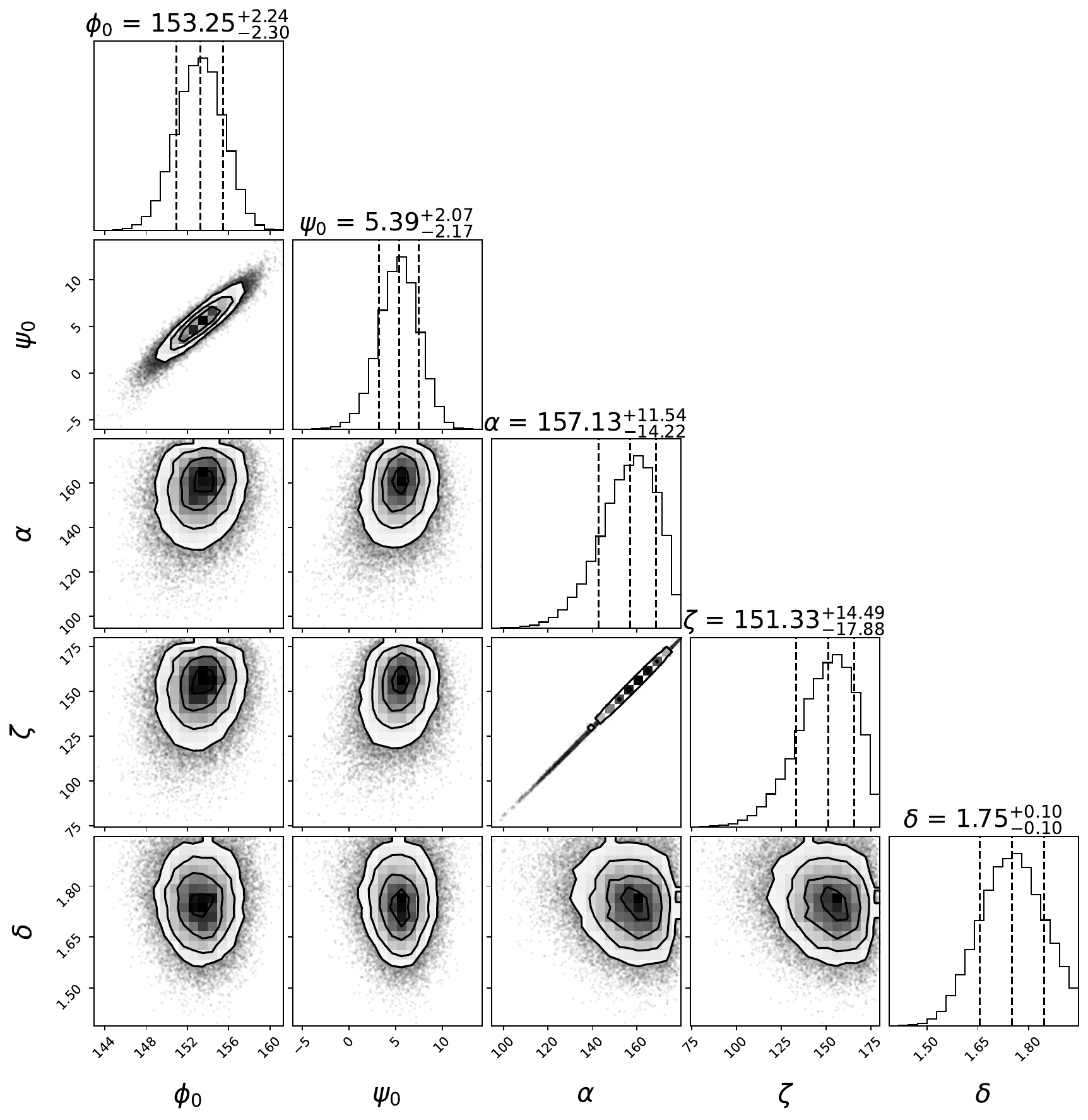}}
 \caption{Posterior distributions of the five parameters in RVM fitting as resulting from MCMC for PSR J0941$-$39. The dashed vertical lines indicate the median and the 16 and 84 percentiles of the distribution, respectively.}
 \label{fig:1rvm}
 \end{figure*}

\subsection{RVM fitting of PSR J0941$-$39}

The PA swings in both the pulsar and RRAT states of PSR J0941$-$39 exhibit S-shapes. We performed RVM~\citep{rc1969} fitting for these two emission states. The PA variation across the pulse phase ($\phi$) can be described by the following equation~\citep{bcw1991}:
\begin{equation}\label{eq1}
\tan(\psi-\psi_{0})=\frac{\sin\alpha\sin(\phi-\phi_{0})}{\sin\zeta\cos\alpha-\cos\zeta\sin\alpha\cos(\phi-\phi_{0})}
\end{equation}
Here, $\zeta=\alpha+\beta$, where $\alpha$ represents the angle between the rotation axis and magnetic axis, while $\beta$ represents the angle between the magnetic axis and observer's line of sight, $\psi_{0}$ and $\phi_{0}$ correspond to the PA and pulse longitude at which inflection point of the PA swings.

We used Markov Chain Monte Carlo (MCMC) fitting~\citep{jk2019} using the python package EMCEE~\citep{fhl+2013} to determine the values of these unknown parameters. The obtained fitting results are represented by the green lines in panel (a) of Figure~\ref{fig:1pa}, and the corner plots are shown in Figure~\ref{fig:1rvm}. The pulsar geometry is characterized by $\alpha$ and $\beta$. We found that both the pulsar and RRAT states exhibit similar values for $\alpha$ and $\zeta$ within the range of uncertainties, implying an unchanged magnetospheric geometry across different states. The approximate 180\,deg value of $\alpha$ suggests that PSR J0941$-$39 is a nearly-aligned rotator.

The absolute height of the emission region with respect to the center of the pulsar can be estimated by measuring the delay ($\delta\phi$) between the steepest gradient point of the PA swings and the profile center~\citep{bcw1991}:
\begin{equation}\label{eq1}
h_{\rm em}=\frac{Pc \delta \phi}{8\pi} 
\end{equation}
Here, $c$ represents the speed of light, and $P$ represents the rotation period of a pulsar. It should be noted that the emission beam for a pulsar may not be filled, or may not be circular and can be compressed in longitude \citep{jk2019mn,dkl+2019}, {which may lead to} an underestimation of the central position of the pulse profile. Therefore, we do not measure the emission height of each emission state. We assume that both profiles in pulsar and RRAT states share identical profile centers. The difference in emission height between these two states can be described as follows: 
\begin{equation}\label{eq1}
h_{\rm diff}=\frac{Pc (\phi_{\rm PSR} - \phi_{\rm RRAT})}{8\pi},
\end{equation}
where $\phi_{\rm PSR}$ and $\phi_{\rm RRAT}$ represent the steepest gradient point of the PA swings for  pulsar and RRAT states, respectively. Taking $\phi_{\rm PSR} = 152.3(3)$\,deg and $\phi_{\rm RRAT} = 153(2)$\,deg, we find that $h_{\rm diff} = 86(281)$\,km. {Considering the large uncertainty}, it is reasonable to conclude that emissions from both pulsar and RRAT states occur at approximately similar heights for PSR J0941$-$39.

\section{DISCUSSION AND CONCLUSIONS}\label{sec:discussion}

{The polarization analysis of pulsars that exhibit transitions between the pulsar state and RRAT state provides informations on the geometric properties and locations of radio emissions from both the pulsar and RRAT.} We have obtained polarization profiles for different states of PSR J0941$-$39 and PSR J1107$-$5907. Our results reveal that the PA swings of PSR J0941$-$39 exhibit smooth variations with a similar S-shape, whereas those of PSR J1107$-$5907 exhibit complicated variations. RVM fitting was carried out for both the pulsar and RRAT states of PSR J0941$-$39, revealing similar values for $\alpha$ and $\zeta$, indicating consistent magnetospheric geometry between the RRAT state and the pulsar state. By calculating the difference in emission height between the pulsar and RRAT states of PSR J0941$-$39, we find that these two states emit at identical heights.

In the emission model of~\citet{rs1975}, the characteristic frequency of radio emission $\nu_{\rm c}\sim \gamma^3 c/\rho_c$~\citep{mgp2000,glm2004,mgm2009} is determined by the Lorentz factor of the radiating plasma $\gamma$, the speed of light $c$, and the underlying radius of curvature of the magnetic field $\rho_c$. The power of radio emission $P \propto F(Q) \gamma^4 c/\rho_c^2$, where $F(Q)$ is associated with plasma properties~\citep{bmm2017}. For J0941$-$39, both pulsar and RRAT emissions occur at a similar altitude, indicating a shared value for $\rho_c$. Therefore, variations in $F(Q)$ are likely responsible for the occurrence of RRATs. The appearance of RRAT state in PSR J0941$-$39 could be attributed to global changes in currents within its magnetosphere.

The emission variations of PSR J0941$-$39 and PSR J1107$-$5907 are reminiscent of the mode changing phenomenon in pulsars, where the average pulse profile abruptly changes between two or more states~\citep{bh1982}. Mode changing is believed to share a common physical origin with pulsar nulling~\citep{wmj2007}. \citet{t2010} proposed that changes in current distribution or magnetospheric geometry of the pulsar could lead to mode changing. Analysis of polarization profiles suggests that emissions from different states of many pulsars originate from the same location within the pulsar magnetosphere and that
the magnetospheric geometry remains constant, e.g.,~\citet{bmm2017,svw+2022,jcb+2022}. It is likely that mode changing is driven by variations in currents within the pulsar magnetosphere~\citep{Lyne2010Sci,bmm2017,syw+2022}, which is similar to the case of PSR J0941$-$39.

By analyzing the polarization profile across a wide bandwidth for different states of these two pulsars,  we found that both the $f_{\rm L}$ and $f_{\rm C}$ of the two pulsars exhibit non-monotonic variations with increasing {frequency. Using} the UWL receiver on the Parkes 64\,m radio telescope, \citet{sjd+2021} presented polarization profiles of 40 bright pulsars and observed a general trend where $f_{\rm L}$ decreases while $f_{\rm C}$ increases with increasing frequency. However, for a specific pulsar, both $f_{\rm L}$ and $f_{\rm C}$ can exhibit intricate variations with frequency. In general, our results are consistent with those reported by~\citet{sjd+2021}.

The spectral index of PSR J0941$-$39 is steeper in its pulsar state compared to its RRAT state, whereas for PSR J1107$-$5907, it is flatter in its pulsar state than in its RRAT state. It remains uncertain whether RRATs possess identical spectral indices to pulsars. A study carried out by~\citet{jvk+2018} investigated the spectral properties of 441 pulsars and obtained a weighted mean spectral index of $-1.60(3)$. Using single pulses, \cite{mtb+2019} measured the spectral index of RRAT J2325$-$0530 as $-2.2(1)$,~\citet{xww+2022} determined that RRAT J0139+3336 has a spectral index of $-3.2(2)$, and \citet{smk2018} measured the spectral indices of three RRATs with values for RRAT J1819$-$1458, RRAT J1317$-$5759, and RRAT J1913+1330 being $-1.1(1)$, $-0.6(1)$, and $-1.2(2)$, respectively. For PSR J0941$-$39 and PSR J1107$-$5907 in their respective RRAT states, the observed spectral indices fall within the known range observed in other RRATs, with values of $-0.58(8)$ and $-0.82(5)$, respectively. Further observations for RRATs using the Parkes UWL receiver~\citep{hmd+2020} would provide more detailed information on the spectral properties.

We also note that PSR J0941$-$39 and PSR J1107$-$5907~\citep{yws+2014} are nearly-aligned rotators,  as well as PSR B0826$-$34~\citep{eam+2012}. It remains unknown whether {nearly-aligned} pulsars exhibit a preference for displaying RRAT-like emissions. Polarization analysis serves as a valuable tool in determining the inclination angle of a pulsar, {however, the polarization properties of RRATs remain poorly explored.} In future investigations, we intend to carry out polarization analysis on a sample of RRATs using the Five-hundred-meter Aperture Spherical radio Telescope (FAST, \citealt{nlj2011}).

\section*{Acknowledgments}
This work is sponsored by the National Natural Science Foundation of China (NSFC) project (No.12288102, 12041303, 12041304, 12273100, 12203092), the Major Science and Technology Program of Xinjiang Uygur Autonomous Region (No. 2022A03013-1, 2022A03013-3, 2022A03013-4), the National Key R\&D Program of China (No. 2022YFC2205201, 2022YFC2205202, 2020SKA0120200), the Natural Science Foundation of Xinjiang Uygur Autonomous Region (No. 2022D01B218, 2022D01B71, 2022D01D85), the West Light Foundation of Chinese Academy of Sciences (No. WLFC 2021-XBQNXZ-027), the 2022 project Xinjiang uygur autonomous region of China for Tianchi talents. The research is supported by the Scientific Instrument Developing Project of the Chinese Academy of Sciences, Grant No. PTYQ2022YZZD01, and partly supported by the Operation, Maintenance and Upgrading Fund for Astronomical Telescopes and Facility Instruments, budgeted from the Ministry of Finance of China (MOF) and administrated by the Chinese Academy of Sciences (CAS). The Parkes radio telescope (Murriyang) is part of the Australia Telescope National Facility that is funded by the Australian Government for operation as a National Facility managed by CSIRO. { We thank the referee for providing valuable suggestions, which greatly improved the paper}.

\software{{\tt DSPSR} \citep{vb2011}, {\tt PSRCHIVE} \citep{hvm2004}}

\bibliography{sample63}{}
\bibliographystyle{aasjournal}

\end{document}